\documentclass{aa} 

\usepackage[varg]{txfonts}
\usepackage{amsmath}
\usepackage{graphicx}
\usepackage{txfonts}
\usepackage{lscape}
\usepackage{amssymb}
\usepackage{mathrsfs}
\usepackage{stmaryrd}
\usepackage{subcaption}
\usepackage{lmodern}
\usepackage{wrapfig}
\usepackage{blindtext}
\usepackage{enumitem}
\usepackage{amssymb}
\usepackage{esvect}
\usepackage{empheq}
\usepackage{cancel}
\usepackage{hhline}
\usepackage{times}
\usepackage{color}
\usepackage{subeqnarray}

\begin{document}

\title{Eigenvectors, Circulation and Linear Instabilities for Planetary Science in 3 Dimensions (ECLIPS3D)}

\titlerunning{ECLIPS3D}
\authorrunning{Debras et al.}

\author{F. Debras \thanks{corresponding author: florian\_debras@hotmail.com} \inst{1,2,3} \and N. Mayne \inst{2} \and I. Baraffe \inst{1,2}
\and T. Goffrey \inst{2} \and J. Thuburn \inst{4}}


\institute{Ecole normale sup\'erieure de Lyon, CRAL, UMR CNRS 5574, 69364 Lyon Cedex 07,  France 
\and 
School of Physics and Astronomy, University of Exeter, Exeter, EX4 4QL, UK
\and
IRAP, Universit\'e de Toulouse, CNRS, UPS, Toulouse, France
\and
College of Engineering, Mathematics and Physical Sciences, University of Exeter, Exeter, EX4 4QF, UK}

\date{}

\abstract
{The study of linear waves and instabilities is necessary to understand the physical evolution of an atmosphere, and can provide physical interpretation of the complex flows found in simulations performed using Global Circulation Models (GCM). In particular, the acceleration of superrotating flow at the equator of hot Jupiters has mostly been studied under several simplifying assumptions, the relaxing of which may impact final results.}
{We develop and benchmark a publicly available algorithm to identify the eigenmodes of an atmosphere around any initial steady state. We also solve for linear steady states indicated to be essential in existing theories of the acceleration of hot Jupiter superrotation. }
{We linearise the hydrodynamical equations of a planetary atmosphere in a steady state with arbitrary velocities and thermal profile. We then discretise the linearised equations on an appropriate staggered grid, and solve for eigenvectors and linear steady solutions with the use of a parallel library for linear algebra: ScaLAPACK. We also implement a posteriori calculation of an energy equation in order to obtain more information on the underlying physics of the mode.}
{Our code is benchmarked against classical wave and instability test cases in multiple geometries (2D, 3D, two layer equivalent depth). The steady linear circulation calculations also reproduce expected results for the atmosphere of hot Jupiters.  We finally show the robustness of our energy equation, and its power to obtain physical insight into the modes.}
{We have developed and benchmarked a code for the study of linear processes in planetary atmospheres, with an arbitrary steady state. The calculation of an a posteriori energy equation provides both increased robustness and physical meaning to the obtained eigenmodes. This code can be applied to various problems, and notably to further study the initial spin up of superrotation of GCM simulations of hot Jupiter.}

\keywords{Hydrodynamics -- Waves -- Instabilities --Planets and satellites: atmospheres -- Methods: numerical}

\maketitle

Accepted in A\&A

\section{Introduction}
 \label{sec:intro}

The study of the influence and propagation of waves in planetary atmospheres is often performed under several simplifications, most notably the assumption of a zero or zonally--symmetric and constant initial zonal flow \citep[e.g.][]{Kasahara2000}, or restriction to a beta--plane solution \citep[e.g.,][]{Lindzen1967}. Such simplifications allow analytical prediction of the key wave mechanisms, and in some cases a complete understanding of their structure \citep[even in the mathematical sense, see][]{Matsuno}. 

Despite the simplifications, such studies have allowed significant insight into atmospheric dynamics, for example, \citet{Wheeler1999} demonstrate that the propagation of waves can be linked to convective motions in Earth's atmosphere, and the resulting description matches analytical theories \citep[e.g.,][]{Vallis06,Holton1992} remarkably well. Baroclinic and barotropic instabilities are also known to have an impact on the circulations of planetary atmospheres \citep[see][]{Williams2003}.

However, for more complex mean flows, or situations where individual terms in the hydrodynamical equations are not clearly dominant analytical treatments rapidly become impractical. Additionally, some wave structures, or modes, are only supported by the full equations, being effectively `filtered' out by the simplifications. Notably, \citet{Wang_Mitchell} numerically identify a Rossby--Kelvin wave mode for a planetary atmosphere that can not be recovered in the quasi--geostrophic equations (see \citet{Gill1980} or \citet{Vallis06} for further details).

The detection and characterisation of a specific class of exoplanets, hot Jupiters, has provided impetus to the study of non-axisymmetrically forced planetary atmospheres. Hot Jupiters are Jovian planets, in short-period orbits close to their host star, and likely have synchronised rotational and orbital speeds, such that a single hemisphere, or day side, faces the host star at all times \citep[see][]{Baraffe2010}. The slow rotation (periods of $\sim$4\,days) and Jovian radii suggest such atmospheres exist in a regime where the Rossby number is of order unity $R_{\rm   o}=\frac{U}{Lf}$, where $U$ is the characteristic flow velocity, $L$ a characteristic length scale and $f$ the Coriolis parameter, meaning rotation is neither dominant, nor--negligible. Observational evidence has indicated the presence of fast zonal ``jets'' (zonally coherent flows) of a few kms$^{-1}$ \citep{Louden2015}. The mechanism for accelerating these zonal flows has been explored by \citet{Showman2011}, building on the linear studies of \citet{Matsuno} and \citet{Gill1980} and using a two--layer equivalent depth approach. 

To further study the linear waves and instabilities present in these atmospheres, we have developed a public code, ECLIPS3D \footnote{\url{https://github.com/fdebras/ECLIPS3D}}(Eigenvectors, Circulation and Linear Instabilities for Planetary Science in 3 Dimensions) which we benchmark in this work. More globally, this code can be used in the study of linear stable or unstable modes within any planetary atmosphere from an arbitrary initial steady state.

We expand upon \citet{Thuburn2002}, who studied propagating wave modes in an axisymmetric atmosphere at rest, to include linear modes in an atmosphere with a steady background flow, in three dimensional spherical coordinates. We detail the structure of ECLIPS3D including the equations solved and the process of obtaining a solution. We detail the different sets of equations implemented (axisymmetric 2D, 3D, two-layer equivalent depth) as well as the time-dependent or independent solutions (waves, instabilities and standing circulation). Finally, we have implemented a posteriori calculation of an energy equation for a given solution. These semi--analytical results allow the verification of the frequency of the modes (and the growth and damping rate for instabilities), as well as isolation of the dominant mechanism providing insight into the physical phenomena driving the instability.

In Section \ref{sec:presentation}, we outline the equations implemented in ECLIPS3D (with the full equations detailed in Appendix \ref{app:full_equations}), and the procedure for solving them, alongside the method of calculating the energy equation. We then benchmark ECLIPS3D against a range of classical calculations of waves, instabilities and circulations in Section \ref{sec:waves}, including a setup similar to \citet{Showman2011}. Finally, we draw conclusions and comment on future developments and applications for ECLIPS3D in Section \ref{sec:conclusion}.

\section{The algorithm}
\label{sec:presentation}

\subsection{ Linearised equations}
\label{ssec:eq}

\citet{Thuburn2002} show that even the simplest atmospheric waves exhibit behaviour that cannot be accurately expressed by separating variables \citep[requiring a height--dependent shift in latitude, see][for details]{Thuburn2002}.
Therefore, in the general case, no assumption can be made on the mathematical expression of the wave regarding spatial coordinates. As our steady state is arbitrary, we will linearise the full equations with no simplification. However, to more easily describe the main capabilities of ECLIPS3D we detail how ECLIPS3D solves the Euler equations, omitting diffusion or viscosity, although
dissipative processes have been implemented (and discussed in the steady circulation case in section \ref{ssec:steady} and Appendix \ref{sapp:steady}). This basic equation set is:
\begin{subequations}
\begin{align}
&\dfrac{D u}{D t}- 
2 \Omega v \sin(\phi)+ 2 \Omega w\cos(\phi)+ \dfrac{1}{\rho r \cos(\phi)} \dfrac{\partial p}{\partial \lambda} 
 \nonumber \\
 &+ \dfrac{uw}{r} - \dfrac{uv\tan(\phi)}{r}= 0  \label{eq:Complete_u}\\ 
\ 
&\dfrac{D {v}}{D t}+
2 \Omega u \sin(\phi) + \dfrac{1}{\rho r} \dfrac{\partial p}{\partial \phi} 
 + \dfrac{vw}{r} + \dfrac{u^{2}\tan(\phi)}{r} = 0 \label{eq:Complete_v}\\  
\
&\dfrac{D {w}}{D t} -
2 \Omega u \cos(\phi) + \dfrac{1}{\rho} \dfrac{\partial p}{\partial r} + g - 
\dfrac{u^{2} + v{2}}{r} = 0 \label{eq:Complete_w}\\
\
&\dfrac{D {\rho}}{D t} + \rho \left( 
\dfrac{1}{r \cos(\phi)} \dfrac{\partial u}{\partial \lambda} +  \dfrac{1}{r \cos(\phi)}  
\dfrac{\partial}{\partial \phi} \left(v \cos(\phi) \right)  \right. \nonumber \\
& \left. +\dfrac{1}{r^{2}} 
\dfrac{\partial}{\partial r} \left( r^{2} w \right)  \right) = 0 
\label{eq:Complete_p}\\
&\dfrac{D \theta}{D t}= \dfrac{\theta}{T} \dfrac{Q}{c_{p}}
\label{eq:Complete_energy}
\end{align}
\begin{align}
&p = \rho R T \label{eq:ideal} \\
&\theta = T \left(\dfrac{p_0}{p}\right)^{\frac{R}{c_{p}}},\label{eq:closure}
\end{align}
\label{eq:Complete_eq}
\end{subequations}
where $u$,$v$ and $w$ are the components of velocity in the longitudinal ($\lambda$), latitudinal ($\phi$) and vertical ($r$) directions, $\rho$ is the density, $p$ the pressure, $T$ the temperature,  $\theta$ the potential temperature, $p_0$ a reference pressure, $R$ is the gas constant divided by mean molecular weight, $c_p$ the heat capacity, $g$ the gravitational acceleration (and is a function of $r$, see Appendix \ref{sapp:3D}), $\Omega$ the rotation rate of the planet, $r$ the radial distance from the centre of the planet, $\lambda$ the longitude, $\phi$ the latitude and finally $Q$ is the heating rate (if present). Equations \eqref{eq:Complete_u} to \eqref{eq:Complete_w} represent momentum conservation, \eqref{eq:Complete_p} mass conservation, \eqref{eq:Complete_energy} conservation of energy, \eqref{eq:ideal} is the equation of state (here an ideal gas) and \eqref{eq:closure} defines potential temperature, closing the set. \citet{Thuburn2002} showed that potential temperature is more appropriate than normal temperature for studies of the linear modes. Finally, $D/Dt$ is the Lagrangian or material derivative and $t$ is time.

Solving for waves or instabilities then requires linearising these equations. We follow the definitions of \citet{Thuburn2002} for the perturbed variables (this scaling comes 
from \citet{Daley1988}), which greatly simplify the equations when the steady state is axisymmetric and at rest. This choice has been made for easier comparison and benchmarking with \citet{Thuburn2002}, but a user of the code can change the implemented equations easily without affecting the method of solution. Namely, we write

\begin{subequations}
\begin{align}
&u'=\rho_{i} \left(u - u_{i} \right)\\
&v'=\rho_{i} \left(v - v_{i} \right)\\
&w'=\rho_{i} \left(w - w_{i} \right)\\
&p'= \left(p-p_{i} \right) \\
&\theta'= \dfrac{g \rho_{i}}{\theta_{i}} \left(\theta - \theta_{i} \right),
\end{align}
\label{eq:perturb_var}
\end{subequations} 
where a prime denotes a linearised variable and an $i$ subscript the initial steady state.

If the heating rate $Q$ is non zero, 
it has to be properly included in the linearised equations. When solving for waves and instabilities, 
we simply linearise $Q$ and include it in the left hand side of the equations. This is detailed in Appendix \ref{sapp:heat}. When looking for steady, linear circulation (and not free or forced waves) $Q(r,\phi,\lambda)$ is specified and considered small enough to only trigger a linear response. A dissipative mechanism must also be added in order to reach a linear steady state. This setup is similar to that of \citet{Showman2011}, which is one of our benchmark cases, and is detailed further in Appendix \ref{sapp:steady}.

\citet{Thuburn2002} considered the linearised equations for the case where the initial atmospheric state is axisymmetric, in hydrostatic balance and at rest. In the more general case linearisation of each of the terms from Eq.(\ref{eq:Complete_eq}) must be completed as shown in Appendix \ref{app:full_equations}, alongside the resulting final equation set Eq.(\ref{final_u}-\ref{final_p}). These final, linearised equations are then discretised and solved within ECLIPS3D as detailed in Section \ref{ssec:method}.
Additionally, we have implemented a two-layer model following \citet{Showman2011}, based on their equations 9 and 10 (linearised versions of which are given in Appendix \ref{sapp:twolayer}). Other equation sets (e.g., shallow water, anelastic, ...) and geometries could be implemented within ECLIPS3D with relative ease if required. 

\subsection{Boundary conditions}
\label{ssec:boundary}

For inviscid flows, there must be no normal flow at the limits of the domain in order to obtain a well posed problem with complete boundary conditions.
Namely, we impose that $v' \cos(\phi)$ tends to zero at the poles, and $w'$ is zero at the top of the atmosphere as a no escape condition. At the inner boundary, we impose a solid boundart with $w'$ equals zero. For hot Jupiters, this requires to the modeled domain to extend to high enough pressures for the atmosphere to reach a quiescent region not involved in the acceleration of superrotation. This inner boundary condition can be easily changed if mass flows or energy transfer with the deep atmosphere need to be considered. If the density of the upper atmosphere is too low, unphysical velocities might arise. In the physical applications of ECLIPS3D so far, we have solved this problem by reducing the extent of the atmosphere but a smoothing of the higher atmosphere could be implemented \citep[as done for example in GCMs, see][]{Mayne2014b,Mayne2014}. 

With the above choice of boundary conditions and implemented equations, the code will only recover
standing waves in the vertical direction. However, \citet{Wu2001} have
expressed the importance of vertical wave propagation in the context of Matsuno-Gill structures \citep{Matsuno,Gill1980}, relevant 
for understanding
Earth's climate (see notably \cite{sarachik_cane_2010}) but also regarding the spin-up of superrotation in hot Jupiters \citep{Showman2011}. Addtionally, a numerical way to mimic evanescent waves is to impose a damping region at the top of the atmosphere, 
as is done e.g., for GCM studies of hot Jupiters (see \citet{Mayne2014b}), that prevents the wave from reflecting 
but allow it to propagate.

In this paper, the boundary conditions have been chosen
for benchmarking, but any user of the code can easily apply different boundary conditions.
Additionally, adapting the equations to include
a damping layer in ECLIPS3D poses no theoretical nore numerical issue. Vertically propagating waves can therefore be recovered
with ECLIPS3D.  

\subsection{Energy equation}
\label{ssec:energy}
Following the method of \citet{Thuburn2002}, we 
calculated an energy equation by combining the linearised Euler equations and integrating them over the whole atmospheric volume with appropriate boundary conditions.We derive this equation in the same context as \citet{Thuburn2002}, with an initially axisymmetric, hydrostatically balanced atmosphere at rest. The general case is shown in Appendix \ref{app:energy}. We assume that the linearised variables $X'$ can be expressed as $X'(r,\phi,\lambda,t)  = X(r,\phi) e^{-\mathrm{i} (\sigma t + m \lambda)}$, where the real part of $\sigma$ is the mode frequency and its imaginary part the growth rate (if nonzero), $X(r,\phi) \in \mathbb{C} $, as explained in \ref{ssec:method}, and $m \in \mathbb{Z}$:    
\begin{flalign}
&\int_{0}^{2\pi}\int_{-\pi/2}^{\pi/2}\int_{a}^{a+H} \left. \bigg[ 2 \sigma \rho_i  E - 
F \left(uw^{\star} + u^{\star}w \right)  \right. 
\nonumber \\
& \left. +f \left( u v^{\star} + u^{\star} v \right)
-\dfrac{m}{r \cos{\phi}} \left( u p^{\star} + u^{\star} p\right) \right.
 \nonumber \\ 
 & \left. -
\left(w \theta^{\star} + w^{\star} \theta \right)  
+\dfrac{1}{r} \left(v^{\star} \dfrac{\partial p}{\partial \phi} - +\dfrac{p^{\star}}{\cos(\phi)} 
\dfrac{\partial}{\partial \phi} \left( v \cos(\phi) \right) \right) \right.
\nonumber \\
& \left.+ w^{\star} \dfrac{g}{c_i^2} p - \dfrac{N_i^2}{g} w p^{\star} \right. 
\nonumber \\
 & +\left.w^{\star} \dfrac{\partial p }{\partial r} 
- \dfrac{p^{\star}}{r^2}\dfrac{\partial}{\partial r}
\left( r^2 w \right)  \right] \dfrac{r^2 \cos(\phi)}{\rho_i}  dr d\phi d \lambda = 0,
\label{eq:energy_int0}
\end{flalign}
where $^{\star}$ denotes the complex conjugate, with $f = 2 \Omega \cos(\phi)$, $F = 2 \Omega \sin(\phi)$, $a$ the radius of the planet, $H$ the height of the top of the atmosphere  and $E = 1/2 \left( (|u|^2 + |v|^2 + |w|^2)/\rho_i + |\theta|^2/\rho_i N^2_i + |p|^2/(\rho_i c_i^2) \right)$ the sum of the kinetic, thermobaric and elastic energies of the perturbation,  with $N_i^2$ and $c_i^2$ the initial buoyancy frequency and sound speed, respectively. Using integration by parts we can express:
\begin{align}
&\sigma = -\dfrac{1}{\int\int\int_V E r^2\cos(\phi) dr d\phi d \lambda} \times \nonumber \\
&\int\int\int_V
\dfrac{1}{\rho_i}\Re \left[f \left( u v^{\star} \right)-F \left(uw^{\star}\right)
- \dfrac{m}{r \cos{\phi}} \left( u p^{\star}\right) -
\left(w \theta^{\star} \right) +  \right. \nonumber \\
&\left.  \dfrac{1}{r} \left(v \dfrac{\partial p^{\star}}{\partial \phi} \right) +
\left(w\dfrac{\partial p^{\star} }{\partial r} \right)
+ \dfrac{g}{c_i^2} \left(w p^{\star} \right)
\right] r^2\cos(\phi) dr d\phi d \lambda,
\label{eq:sigma_energy_int}
\end{align}
with V the volume. This is possible only if $ \int\int\int_V E r^2\cos(\phi) dr d\phi d \lambda \neq 0 $ which is the case when $N_i^2 > 0$. For this work we assume a stably stratified atmosphere ($N_i^2 > 0$). As stated by \citet{Thuburn2002}, Eq.(\ref{eq:sigma_energy_int}) shows that $\sigma$ can only be real in this case and no instability can grow around an atmosphere at rest with no heating and the boundary conditions we have described. 

Once the variables $u', v', ...$ are known Eq.\eqref{eq:sigma_energy_int} can then be integrated numerically in ECLIPS3D and be used to verify the obtained frequency and identify the dominant physical processes (e.g., in Section \ref{ssec:thuburn} we show that an acoustic wave is largely dominated by the terms involving the pressure, whereas a Rossby wave is dominated by the $f$ and $F$ terms). 

When using the code, the use of this a posteriori energy equation is therefore a powerful tool both for diagnosis of the dominant physical mechanism, as well as a validation of the numerical results. It requires an interpolation of the output variables and their derivative on a common grid, which in the current version is performed with linear interpolation, but the consistency of the results with the energy equation confirm that a more sophisticated interpolation would not change the physical interpretation provided by this energy equation (see notably section \ref{ssec:wang}).
\subsection{Method of solution}
\label{ssec:method}
The derived linearised equations of motion can be expressed as:
\begin{equation}
D\left( \begin{matrix}
u' \\
v' \\
w' \\
p' \\
\theta'
\end{matrix} \right) = 0,
\label{eq:matrix}
\end{equation}
where $D$ is a differential linear operator. If we introduce the operator $A$ being:
\begin{equation}
A = \left( \begin{matrix}
\dfrac{\partial}{\partial t} & 0 & 0 & 0 & 0 \\
0 & \dfrac{\partial}{\partial t} & 0 & 0 & 0 \\
0 &  0 & \dfrac{\partial}{\partial t} &0 & 0 \\
0 &  0 & 0 &\dfrac{\partial}{\partial t} & 0 \\
0 &  0 & 0 & 0 & \dfrac{\partial}{\partial t}
\end{matrix} \right).
\label{eq:A}
\end{equation}
it is clear that $A$ commutes with $D$ as the latter is only time--dependent through the $\partial/ \partial t$ terms. Therefore, the vector sub--spaces of $D$ remain stable upon application of $A$. As $A$ is diagonal the kernel of $D$ can be decomposed on the eigenvectors $e_{\sigma}$ of $A$. Such eigenvectors are well known: $e_{\sigma} \propto e^{ -i\sigma t}$ where 
$\sigma \in \mathbb{C}$ (the $-\mathrm{i}$ term is just the convention we choose. The real part of $\sigma$
is therefore the frequency and the imaginary part the growth rate). 
Therefore, coupled with appropriate boundary conditions, we can then solve 
Eq.(\ref{eq:matrix}) by decomposing them as:
\begin{equation}
\left( \begin{matrix}
u'(t, r, \phi, \lambda) \\
v'(t, r, \phi, \lambda) \\
w'(t, r, \phi, \lambda) \\
p'(t, r, \phi, \lambda) \\
\theta'(t, r, \phi, \lambda)
\end{matrix} \right) = \sum_{\sigma}
\left( \begin{matrix}
\hat{u}(r, \phi, \lambda) \\
\hat{v}(r, \phi, \lambda) \\
\hat{w}(r, \phi, \lambda) \\
\hat{p}(r, \phi, \lambda) \\
\hat{\theta}(r, \phi, \lambda) 
\end{matrix} \right) \exp^{-i \sigma t},
\end{equation}
with $\hat{u},\hat{v},\hat{w},\hat{p},\hat{\theta} \in \mathbb{C}$ and remembering that the evolution of the perturbed quantity is then the real part of the above expression. The actual solution is an infinite sum over all $\sigma$ but the eigenmodes, and therefore the waves, are the individual projections to a single value. It is worth noting that, a priori, $\sigma$ could take continuous values. For an atmosphere initially at rest, \citet{Matsuno} show that only discrete values are solutions, but on the other hand baroclinic waves exhibits a continuous range of frequencies (see e.g., \citet{Charney1947})

\subsubsection{Time--dependent solution}
\label{ssec:method_dt}
From the discussion above, we can re-write our equations as a complex eigenvalue--eigenvector problem: 
\begin{equation}
B\left( \begin{matrix}
u' \\
v' \\
w' \\
p' \\
\theta'
\end{matrix} \right) = i \sigma 
\left( \begin{matrix}
u' \\
v' \\
w' \\
p' \\
\theta'
\end{matrix} \right ).
\label{eq:B}
\end{equation}
(with $B = D-A$). 

\begin{figure}[ht!]
\centering
\includegraphics[width=\linewidth]{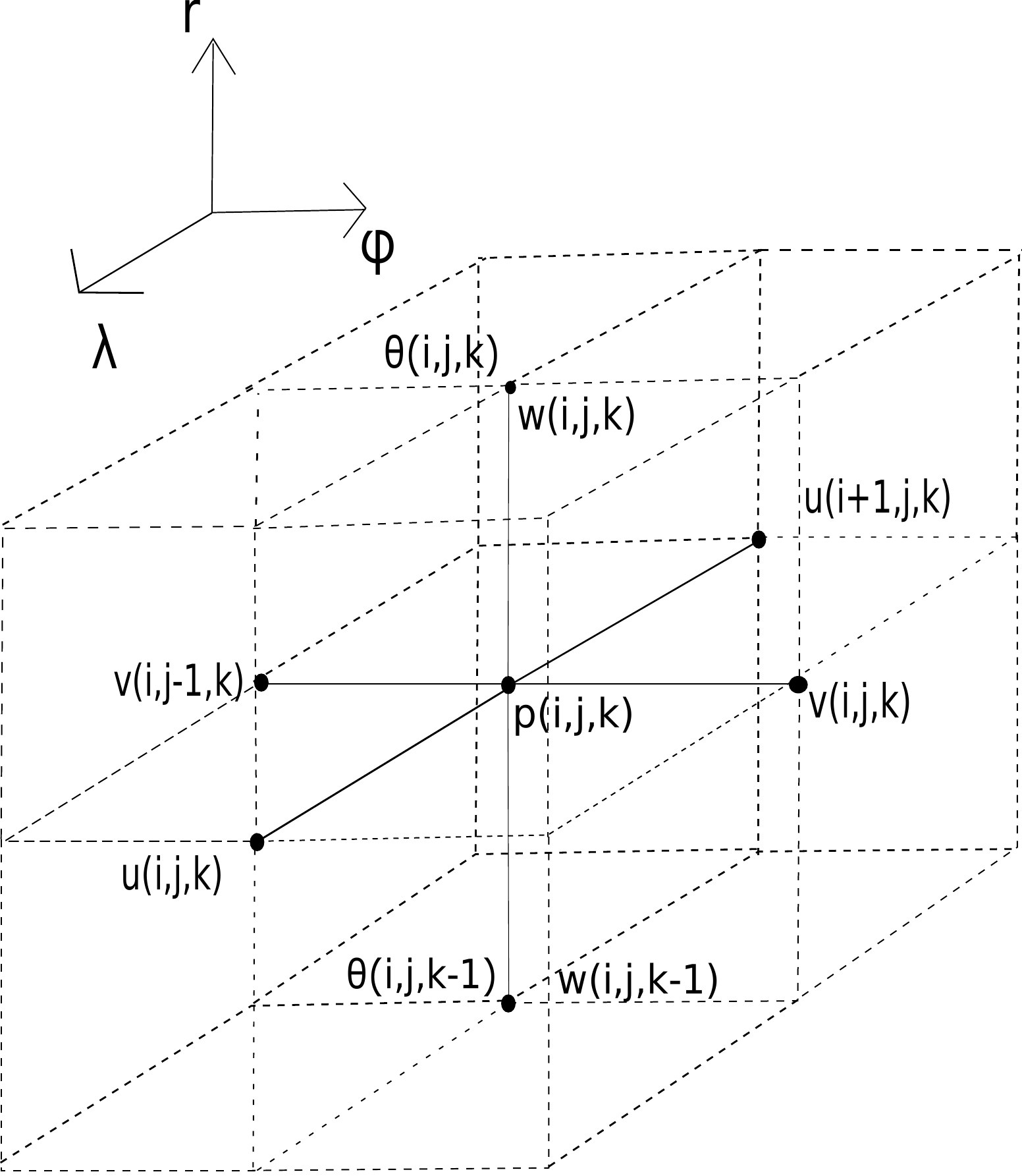}
\caption{Figure showing a cell of the 3D staggered grid adopted in ECLIPS3D, based on \citet{Thuburn2002}. $i$ discretises the 
longitudinal variable $\lambda$, $j$ the latitude $\phi$ and $k$ the radial variable $r$. $u$ and $v$ are staggered in latitude, with $v$ running from the south to north pole and thereby having an additional latitude point. $p$ is staggered in height with $w$ and $\theta$ with the latter two variables running from the bottom to the top of the atmosphere resulting in an additional height point.}
\label{fig:staggered}
\end{figure}

Eq.\eqref{eq:B} can become difficult or even impossible to solve analytically. However, \citet{Thuburn2002} discretise this equation using a staggered grid of points, turning the analytical matrix $B$ into a finite numerical matrix. This allows spatial derivatives to be calculated using finite differences. The staggered grid was selected carefully for precision and stability by \citet{Thuburn2002}, where the 2D axisymmetric version is presented. We adopt a staggered grid in ECLIPS3D, shown in Figure \ref{fig:staggered}, which resembles the one used in \citet{Thuburn2002} but adapted to 3D and with different staggering of the $u$ and $v$ variables at the poles to simplify the boundary condition.

We have $N_{tot} = N_{\lambda}(2 N_{\phi}N_r+(N_{\phi}+1)N_r + 2N_{\phi}(N_r+1))$ points in our grid, with $N_{\lambda}$, $N_{\phi}$ and $N_r$ being the number of points in each coordinate, meaning the matrix $B$ will be of size $(5*N_{tot})^2$ as there are $5$ inter--dependent variables. However, each variable at a given point only depends on the values of all variables over the closest points in the grid. Therefore, $B$ is an extremely sparse matrix.

Once the matrix $B$ is filled with discretised values from Eq.\eqref{eq:B}, at the staggered grid points we must find the eigenvectors of this matrix. ECLIPS3D uses the ScaLAPACK\footnote{\url{http://www.netlib.org/scalapack/}} library for parallel linear algebra \citep{slug}. To express the eigenvectors of a complex matrix we first calculate the upper Hessenberg form of the matrix, then find the Schur decomposition before identifying the eigenvectors themselves \footnote{Handled by ScaLAPACK routines PZGEHRD, PZLAHQR and PZTREVC, respectively.}. Finally, the eigenvectors are returned to their original form via multiplication with the matrix of transformation. 

However, this process of solving for the eigenvectors yields an eigenvector for each row in the matrix. From this set we must select those of interest, representative of physical modes in the atmosphere in question. To identify the interesting eigenvectors we employ two methods. Firstly, for instabilities with positive exponential growth rates, we assume that the modes that will lead the dynamical instability have the highest growth rate, and only select the fastest growing modes. For the case of no instability we first filter out modes arising from numerical errors (e.g., extreme values at the poles), and then manually select modes from the solution set. Analytical expectations then determine the modes of interest. Globally, this selection process needs to be performed thoroughly and based on analytical expectations of the modes to look for. In the current version of the code, the selection is performed indepently of the matrix calculation, and therefore allows to isolate different eigenvectors in a single ECLIPS3D run. 

\subsubsection{Time--independent solution}
\label{ssec:method_nodt}
For the case of a steady circulation, without time--dependence and with constant heating rate Eq.\eqref{eq:matrix} can be expressed as:
\begin{equation}
C * \left( \begin{matrix}
u' \\
v' \\
w' \\
p' \\
\theta'
\end{matrix} \right) =  \left(\begin{matrix}
0 \\
0 \\
0 \\
\gamma R \rho_i  \dfrac{Q_i}{c_p} \\
\dfrac{g \rho_i}{T_i} \dfrac{Q_i}{c_p}
\end{matrix} \right),
\label{eq:C}
\end{equation}
where $C$ is similar to $B$ in Eq.\eqref{eq:matrix} with the inclusion of a drag term if required (see Appendix \ref{sapp:steady}). Solving this problem is much easier than the time--dependent case, as we just need to express $C$ on the staggered grid and invert it to obtain the unique solution to these equations.

\subsection{ECLIPS3D Resolution}

In order to achieve the highest possible resolution, we have implemented two versions of ECLIPS3D: one that solves for the whole eigenvector spectrum and another one dedicated to solving a reduced number of selected eigenvectors. 

For the case of numerically solving for all potential eigenvectors the computational expense (both in computation time and memory) increases steeply with the number of points in the matrix, and can rapidly become inhibitive limiting the resolution. Specifically, within our computational framework, ECLIPS3D can be used to solve 2D problems (axisymmetric, two layer or barotropic equations, see next section) in cases with up to $100\times100$ points within a day of real time. The efficient parallelisation of eigenvector calculations is still an active area of research in the computer science community, and increasing the number of processors does not significantly accelerate calculations of this type. Higher resolution problems therefore take much longer, with our benchmarking tests suggesting the time taken scales with the number of points as roughly $N_{tot}^2$ or $N_{tot}^3$. Additionally, eventually with increasing numbers of processors, the communication between the processors becomes the primary overhead or limitation in the calculation, whereas using too few processors leads to saturation of the available memory (see the ScaLAPACK documentation for details). These computational limitations are amplified in 3D, where calculations at resolutions of $25\times20\times20$ points, on 64 processors require around four days. We are currently working on adapting ECLIPS3D to employ sparse matrix libraries (as dicussed the matrix we are solving is sparse) and are following the developments in computer science research regarding eigenvector calculations. 

However, by solving for a reduce set of selected eigenvectors we can commensurately increase the resolution of the ECLIPS3D setup, whilst retaining a similar computational expense to the case where the target eigenvectors are not restricted. This approach can be taken when there are some existing or prior constraints on the frequency and/or growth rate of the eigenvectors, for example in the case of an instability where one requires the fastest growing mode. Solving for 10 eigenvectors with a resolution of $25\times20\times20$ points takes less than two hours to converge on 64 processors, although these calculations are still limited by the phyiscal memory available on the processors. As mentioned, we are working on implementing sparse matrix solving libraries in ECLIPS3D which will overcome this limitation. 

Globally, the search for particular waves can be optimised by the combination of both a complete eigenvector, and specified or restricted eigenvector setup of ECLIPS3D. First, one would calculate a whole spectrum of low resolution eigenvectors and identify the most interesting ones, before studying them in much higher resolution with the faster version of the code. However, for the time-independent solution, hence matrix inversion, the resolution can be much higher because inverting a matrix is a well parallelised and efficient process,. For example, for $25\times25\times25$ points the matrix inversion takes less than an hour on 16 processors.

Finally, we stress that the symmetries of the problem can allow us to restrict our study to only one hemisphere, doubling the effective resolution with the same number of points. In this paper for example, the modes we present are always symmetric about the equator, but ECLIPS3D can solve for both symmetric and antisymmetric modes at the equator.

\section{Benchmarking}
\label{sec:waves}

We have applied ECLIPS3D to five well studied cases from the literature to benchmark the code. These tests are explained in this section. First we reproduce the results of \citet{Thuburn2002} for a simple, hydrostatically balanced and zonally symmetric atmosphere at rest (Section \ref{ssec:thuburn}). This is followed by the case of an unstable jet providing a steady initial circulation as introduced by \citet{Wang_Mitchell} (Section \ref{ssec:wang}). We also present results for the baroclinic instability test of \citet{Jablo2006,Ullrich2014} (Section \ref{ssec:Ullrich}).
In order to implement longitudinal variation in the steady state, we study the stability of Rossby-Haurwitz waves \citep{Haurwitz1940}, 
as done in \cite{Thuburn2000}. Finally, ECLIPS3D is applied to the case of a linear steady state circulation with atmospheric drag following \citet{Komacek2016} (Section \ref{ssec:steady}).

\subsection{Initial atmospheric rest state}
\label{ssec:thuburn}

\begin{figure*}[ht!]
\centering
  \includegraphics[width=\linewidth]{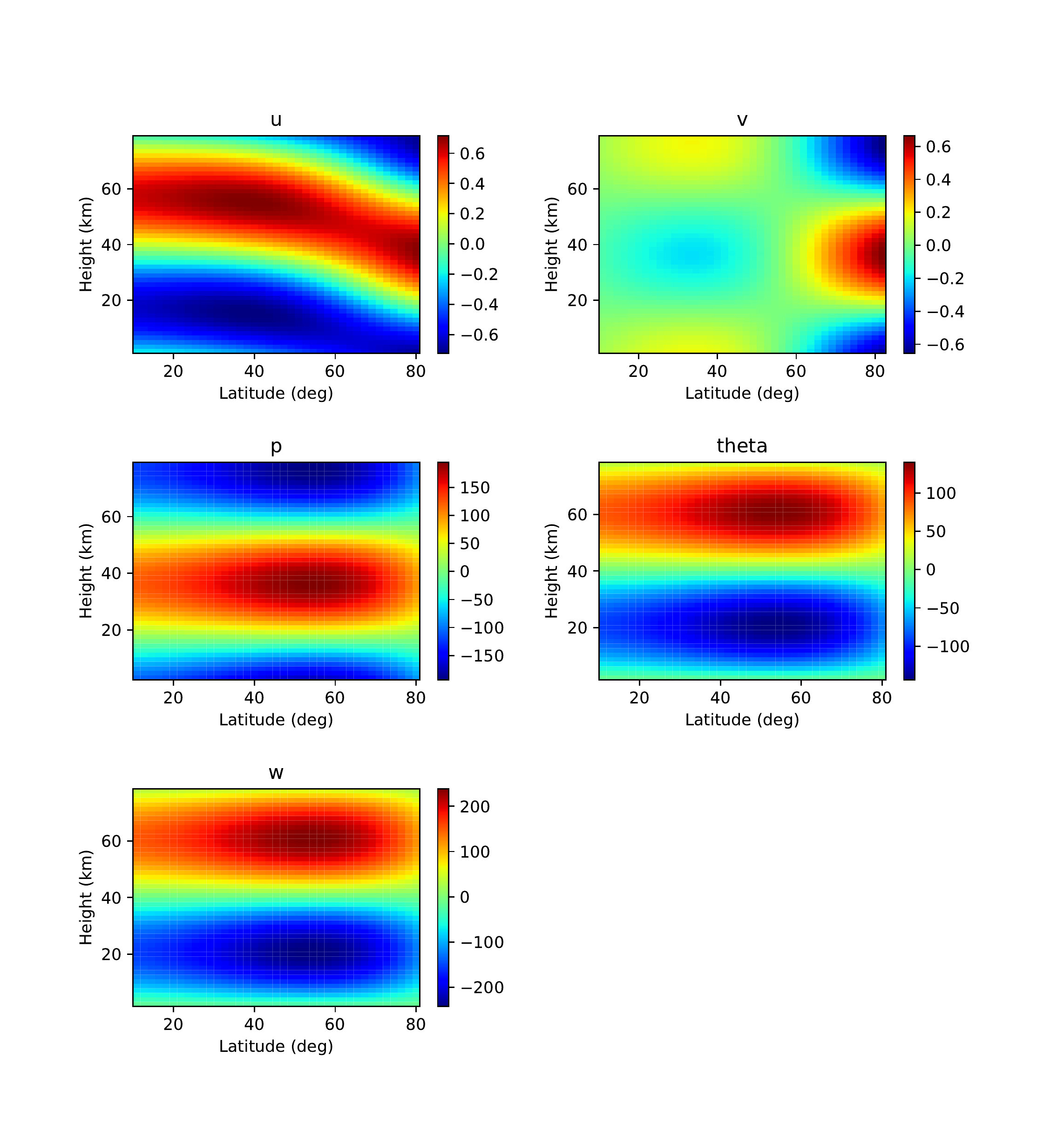}
\caption{Values of the five perturbed variables $u'$,$v'$,$p'$,$w'$ and
$\theta'$ obtained with ECLIPS3D for an acoustic wave with longitudinal wave number 1, with units proportional to their influence on the energy of the wave (as our solutions are from linear theory all values are defined relative to an unknown proportionality value). This mode is to be compared to Figure 2 of \citet{Thuburn2002}}
  \label{fig:Thuburn_acoustic}
\end{figure*}

\begin{figure*}[ht!]
\centering
  \includegraphics[width=\linewidth]{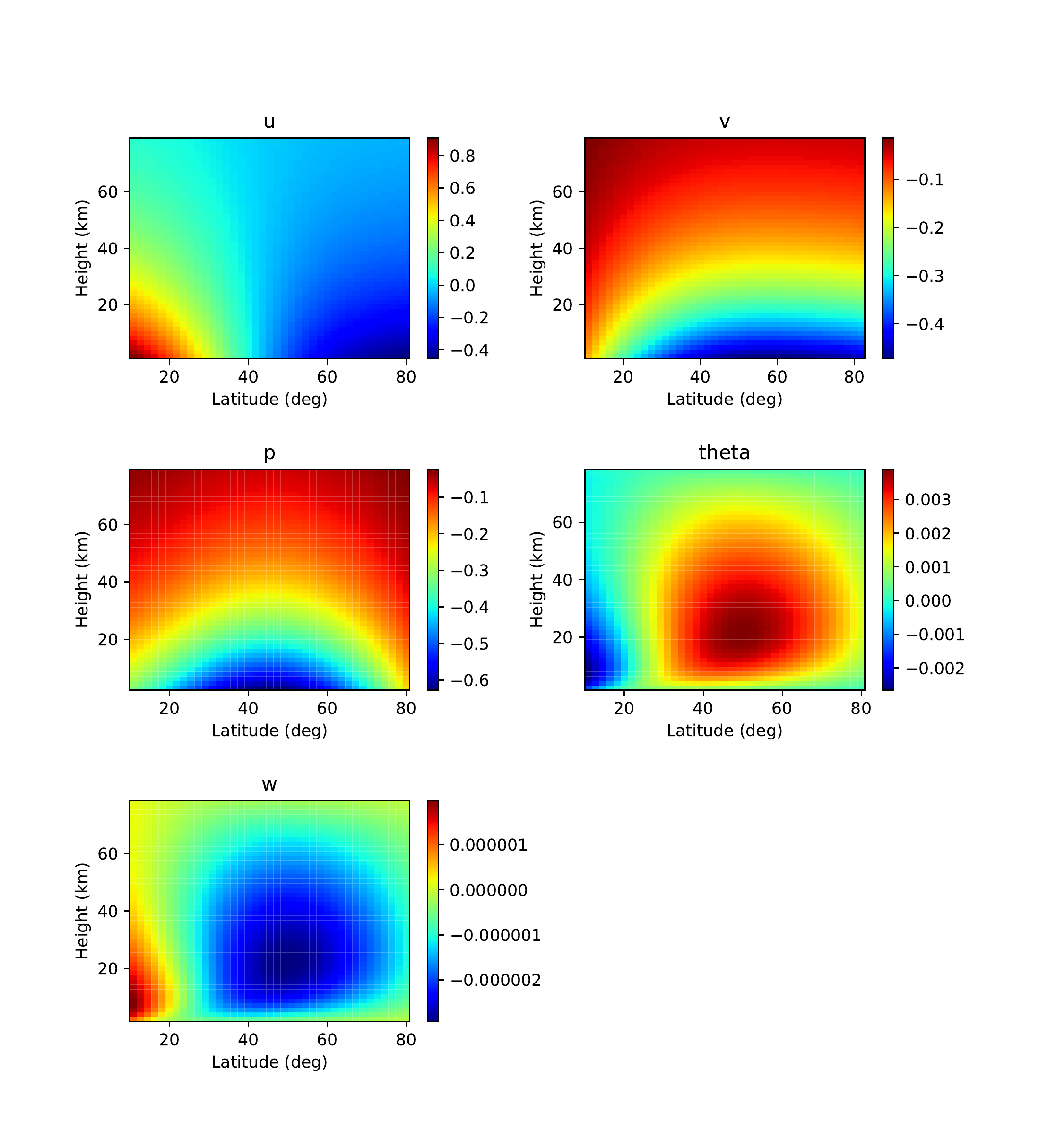}
  \caption{Same as Figure \ref{fig:Thuburn_acoustic} but for a Rossby wave, to be compared
  with Figure 1 of \citet{Thuburn2002}.}
  \label{fig:Thuburn_Rossby}
\end{figure*}

\begin{table*}[ht!]
\begin{center}
\begin{tabular}{ccccccc}
\hhline{=======}
Mode & Acoustic & Acoustic & Gravity & Rossby & Rossby & Kelvin \\
\hline
Thuburn & $3.27\times 10^{-2}$ & $2.87\times 10^{-4}$ & $1.88\times 10^{-4}$ & $-1.46\times 10^{-5}$ & $-3.07\times 10^{-6}$ & $3.14\times 10^{-5}$ \\
ECLIPS3D & $3.28\times 10^{-2}$ & $2.87\times 10^{-4}$ & $1.88\times 10^{-4}$ & $-1.46\times 10^{-5}$ & $-3.02\times 10^{-6}$ & $3.08\times 10^{-5}$ \\
Energy equation & $3.37\times 10^{-2}$ & $2.86\times 10^{-4}$ & $1.88\times 10^{-4}$ & $-1.46\times 10^{-5}$ & $-3.08\times 10^{-6}$ & $3.00\times 10^{-5}$ \\
\hline
\end{tabular}
\caption{Comparison between the frequencies obtained for a sample of different types of waves (see Section \ref{ssec:thuburn}) presented in \citet{Thuburn2002} and those identified in this work using ECLIPS3D. The semi-analytical values from the a posteriori energy equation are also given. All these modes have a longitudinal wavenumber $m=1$.}
\label{tab:comparison_thuburn}
\end{center}
\end{table*}

We first apply the 2D, axisymmetric version of ECLIPS3D to solve for the eigenmodes of an initially axisymmetric, isothermal and hydrostatically balanced atmosphere at rest following \citet{Thuburn2002}. Namely, the atmosphere is $80$km heigh, with the bottom boundary at the Earth radius $a=6371$ km, the temperature is $T = 250$K corresponding to $N^2 = 3.83 \times 10^{-4} \mathrm{s}^{-2}$. The value of the other parameters are $R = 287.05 \mathrm{J.kg}^{-1}\mathrm{K}^{-1}$, $c_p = 1005.0 \mathrm{J.kg}^{-1}\mathrm{K}^{-1}$, $\Omega = 7.292 \times 10^{-5} s^{-1}$ and $g = 9.8062 \mathrm{m.s}^{-2}$. This version of ECLIPS3D needs to assume an integer wavenumber $m$ in longitude, as in \citet{Thuburn2002}. Table \ref{tab:comparison_thuburn} shows the frequencies of the modes from both this study and that of \citet{Thuburn2002} revealing agreement better than 3\%, the discrepancies are due to slightly different initialisations and grid staggering. When matching their setup exactly we return matching results to within machine precision. Additionally, our resulting eigenfunctions have the same shape in height and latitude and global values as those found in \citet{Thuburn2002}. For example, we isolate and present both an acoustic and Rossby wave recovered by ECLIPS3D in Figures \ref{fig:Thuburn_acoustic} and \ref{fig:Thuburn_Rossby}, to be compared to Figures 1 and 2 of \citet{Thuburn2002}. The acoustic mode shows a vertical compression mode, with few energy in the horizontal velocities and an opposite phase between the pressure and vertical velocity perturbations. The tilt in the zonal velocities close to the pole confirms the impossibility to obtain solutions with separate functions in the latitudinal and vertical directiosn in spherical geometry, as noted by \citet{Thuburn2002}. The Rossby modes on the other hand is dominated by pressure and horizontal velocity perturbation, and if projected on a latitude-longitude plane we would recover the rotating winds around pressure maxima or minima in the mid latitudes. Once again the shpae of the waves clearly confirms the impossibility to separate variables.

The difference in the forcing mechanisms between acoustic and Rossby waves, namely the pressure gradient and Coriolis force, respectively, mean we expect their global features to differ. The calculations from our energy equation, shown in Table \ref{tab:comparison_thuburn}, are in excellent agreement with the obtained numerical frequencies. These calculations also allow one to investigate the restoring force. For example, for the Rossby waves the $f$ and $F$f terms of Eq.\eqref{eq:sigma_energy_int} account for $90 \%$ of the value of $\sigma$, whereas they are negligible compared to the terms involving the pressure, and its derivative, for the acoustic waves.

\begin{figure*}[ht!]
  \centering
  \includegraphics[width=\linewidth,height = 12cm]{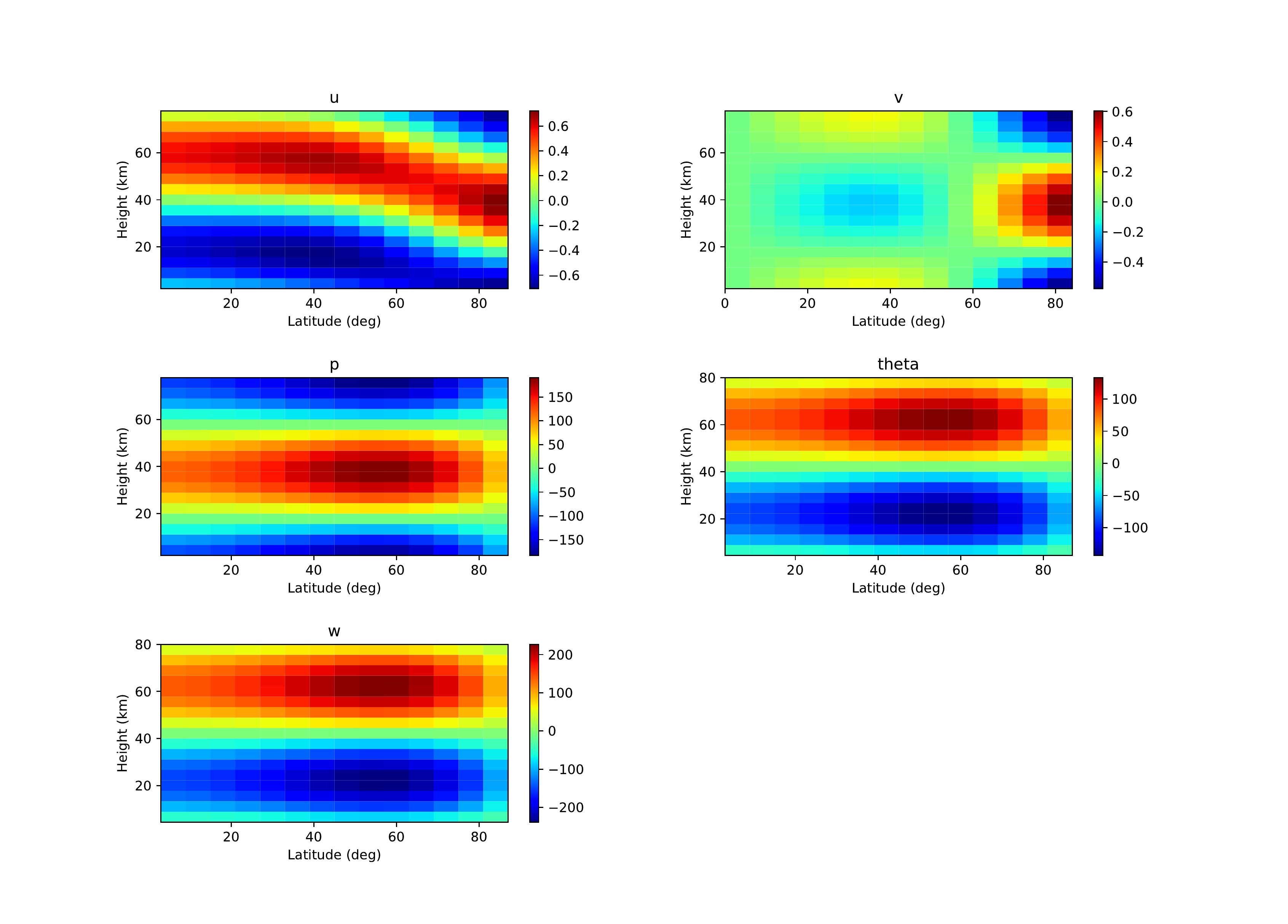}
  \caption{Same as Figure \ref{fig:Thuburn_acoustic} but from the 3D version.}
  \label{fig:3D}
\end{figure*}

Although this problem is axisymmetric, it can be used to test the 3D version of ECLIPS3D. In Figure \ref{fig:3D} we present a single mode from the 3D case, to be compared to Figure \ref{fig:Thuburn_acoustic}. For this mode (and the other modes not presented explicitly here) we obtain the same height and latitude behaviour. For the additional dimension, longitude, we recover oscillatory modes with an arbitrary integer number, $2m$, of zeros in longitude (corresponding to a wavenumber $m$). The frequency of the obtained modes again, as with the 2D version of ECLIPS3D, match those of \citet{Thuburn2002} to better than $\sim 3\%$ (in this case errors are also introduced by the discretisation in longitude).

\subsection{Unstable jet}
\label{ssec:wang}

\begin{figure}[ht!]
  \centering
  \includegraphics[width=\linewidth]{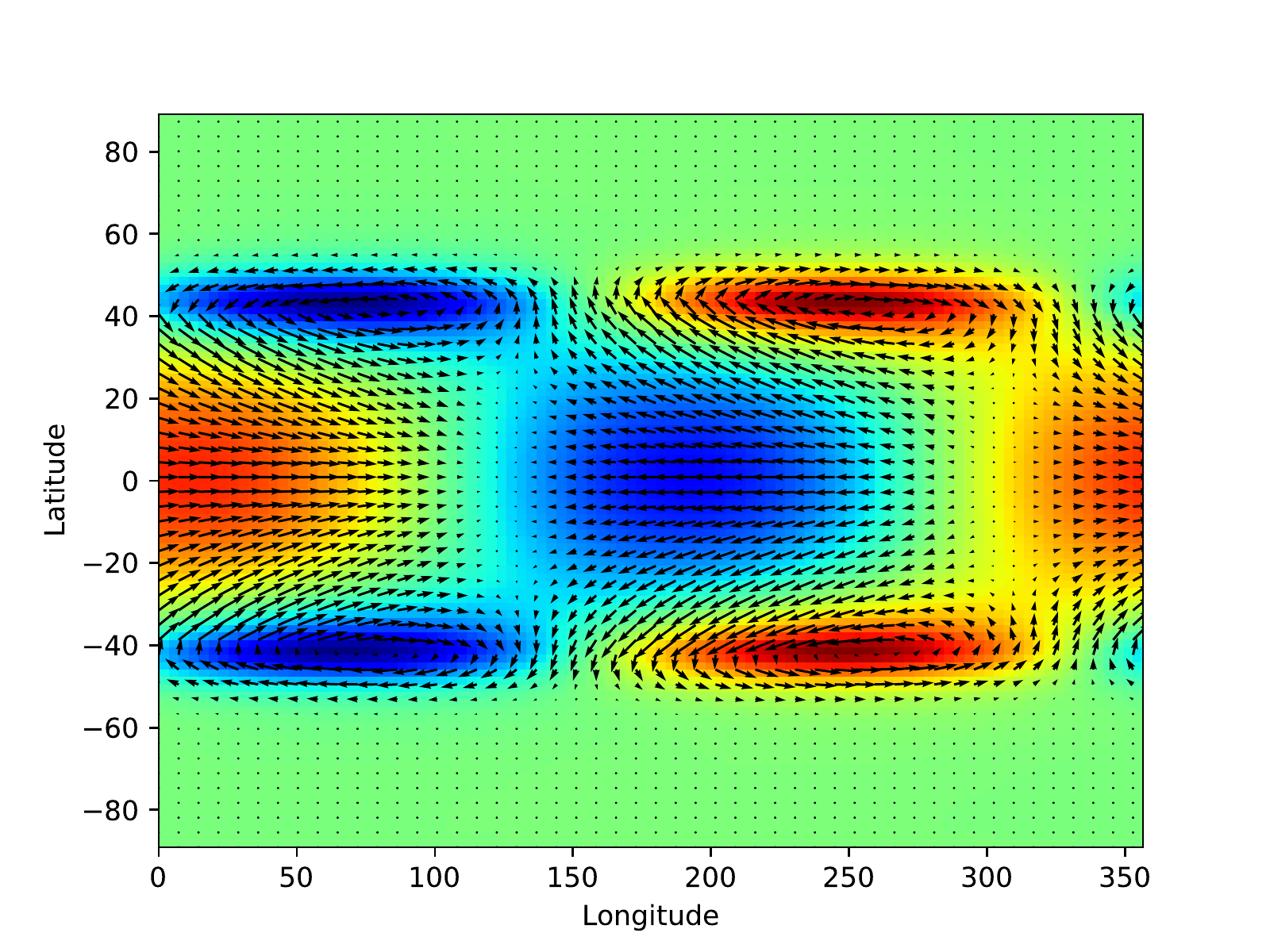}
\caption{Pressure (colour scale) and wind (vector arrows) for the most unstable mode obtained with ECLIPS3D from the setup of \cite{Wang_Mitchell}, which is to be compared with their Figure 1a. }
\label{fig:Wang_mitchell_compare}
\end{figure}

The next benchmark case is an initialstate which includes an initial velocity field. Here we follow \citet{Wang_Mitchell} who identified an exponentially growing linear mode, bringing eastward momentum to the equator, under axisymmetric forcing. This study essentially identifies unstable modes in an atmosphere similar to \citet{Thuburn2002} but including a mid--latitude unstable jet. 
Namely, the initial velocity is controlled by a given latitude $\phi_0$ through:

\begin{align}
&u_i(\phi) = \dfrac{\Omega a \sin^2(\phi)}{\cos (\phi)} \, \text{ for } |\phi| \le \phi_0, \nonumber \\
&u_i(\phi) = \dfrac{\Omega a \sin^2(\phi)}{\cos (\phi)} e^{-\alpha(|\phi|-\phi_0)^2} \, \text{ for } |\phi| > \phi_0.
\end{align}
where $\alpha$ controls the decay of the velocity field towards the pole. The value of $\alpha$ is not given in \citet{Wang_Mitchell}, here we chose $\alpha = 50$ which mimics the shape of their initial velocity field in their Figure 1. 


\citet{Wang_Mitchell} identify two instabilities, firstly a well--known baroclinic instability (such as studied in Section \ref{ssec:Ullrich}), and secondly a new instability not captured by analytical treatments under the $\beta$--plane approximation. This new mode results in the convergence of eastward momentum at the equator, and is related to the Rossby and Froude numbers. \citet{Wang_Mitchell} term this new instability the Rossby--Kelvin instability as it emerges from interaction between the mid--latitude Rossby waves and the Kelvin wave (an equatorially confined gravity wave with zero meridional velocity). In Figure \ref{fig:Wang_mitchell_compare} we show the characteristics of the mode identified by our own study using the 2D axisymmetric ECLIPS3D, which is to be compared with Figure 1a of \citet{Wang_Mitchell}. Figure \ref{fig:Wang_mitchell_compare} demonstrates the excellent agreement of the structure of the mode found using both ECLIPS3D and that of \citet{Wang_Mitchell}.

\begin{figure*}[ht!]
\begin{subfigure}{.5\textwidth}
  \centering
  \includegraphics[width=\linewidth]{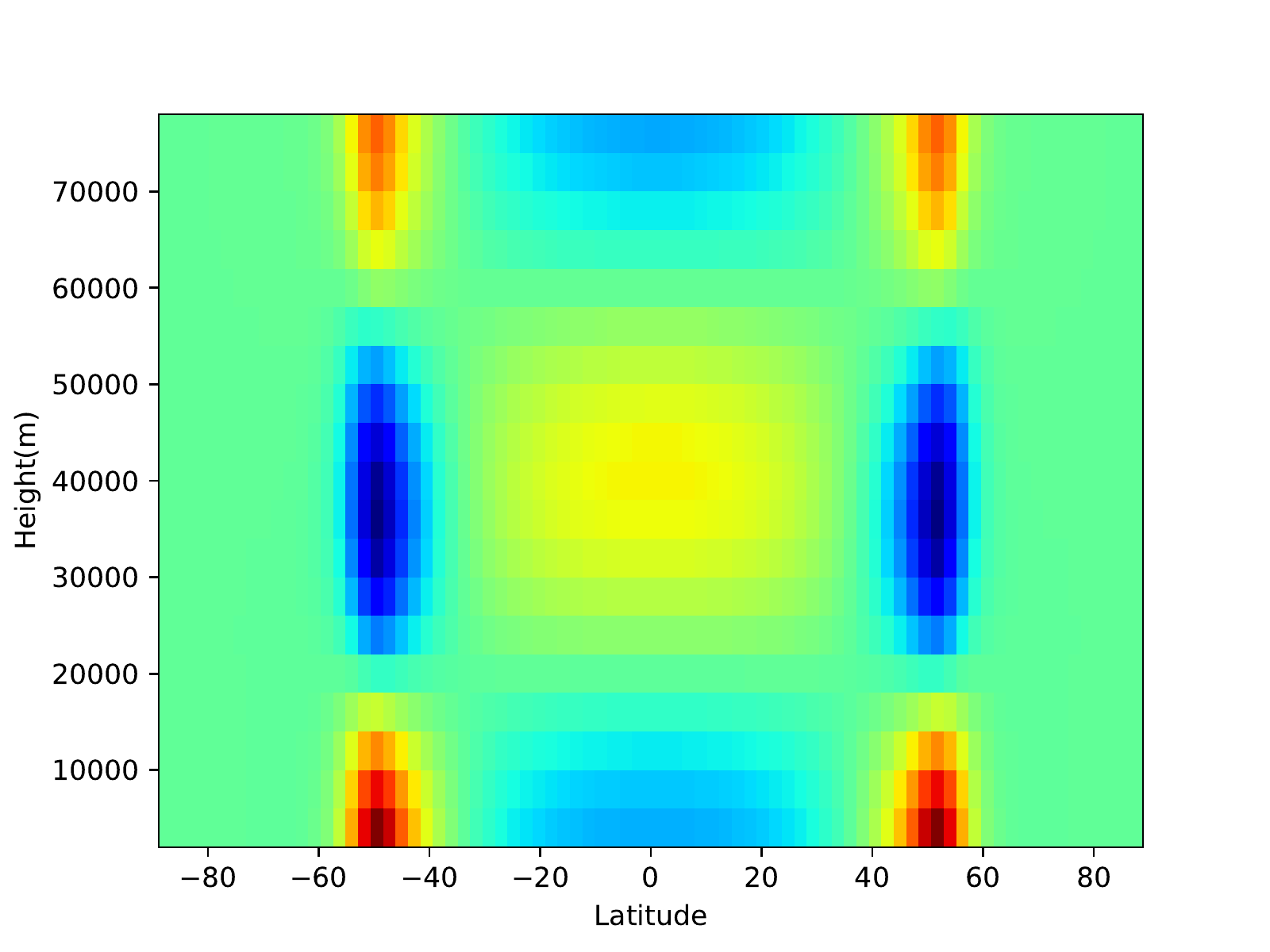}
  \caption{}
  \label{Wang_mitchell_fig_50}
\end{subfigure}%
\begin{subfigure}{.5\textwidth}
  \centering
  \includegraphics[width=\linewidth]{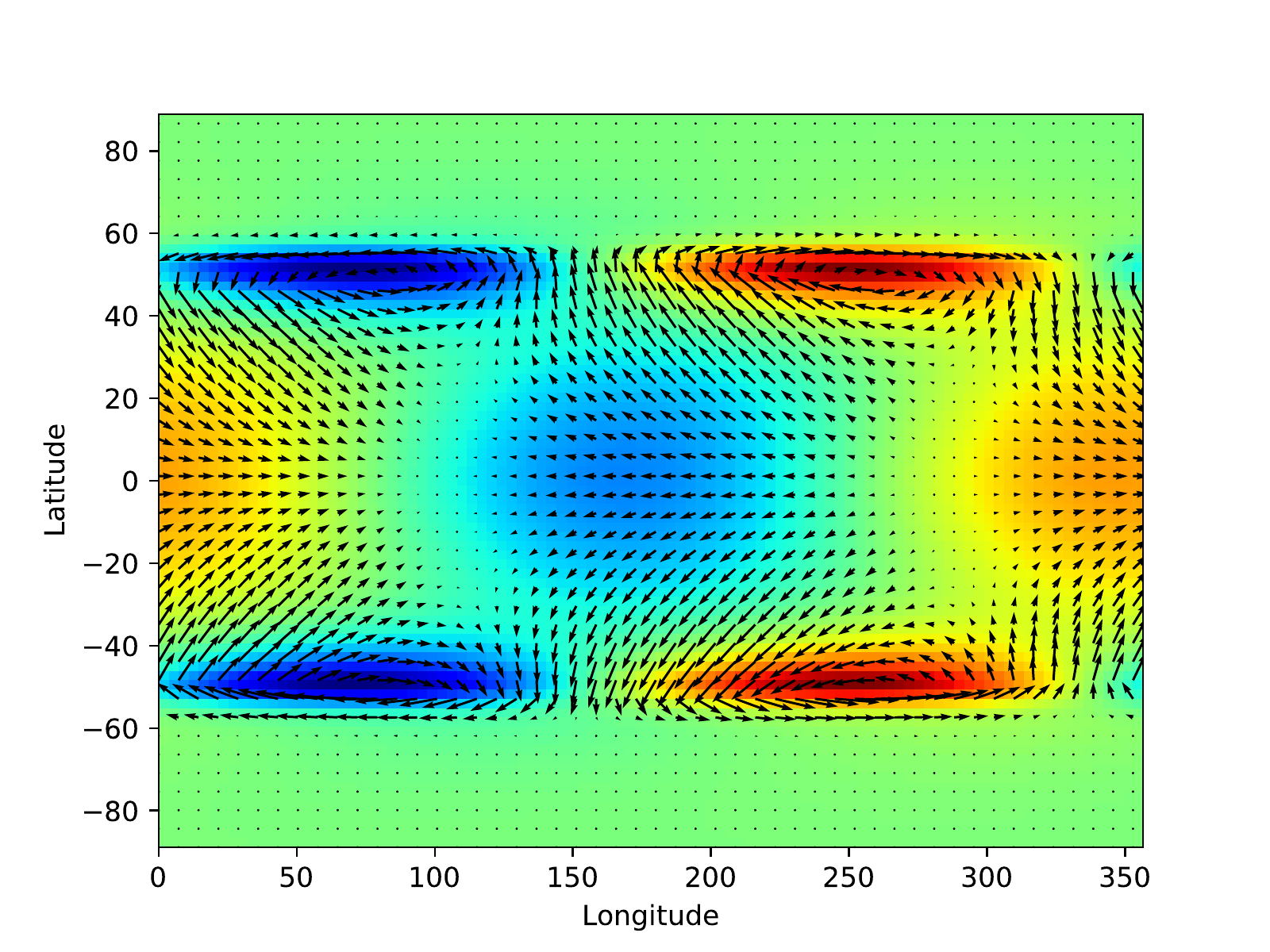}
  \caption{}
  \label{Wang_mitchell_fig_50_ll}
\end{subfigure} \\
\begin{subfigure}{.5\textwidth}
  \centering
  \includegraphics[width=\linewidth]{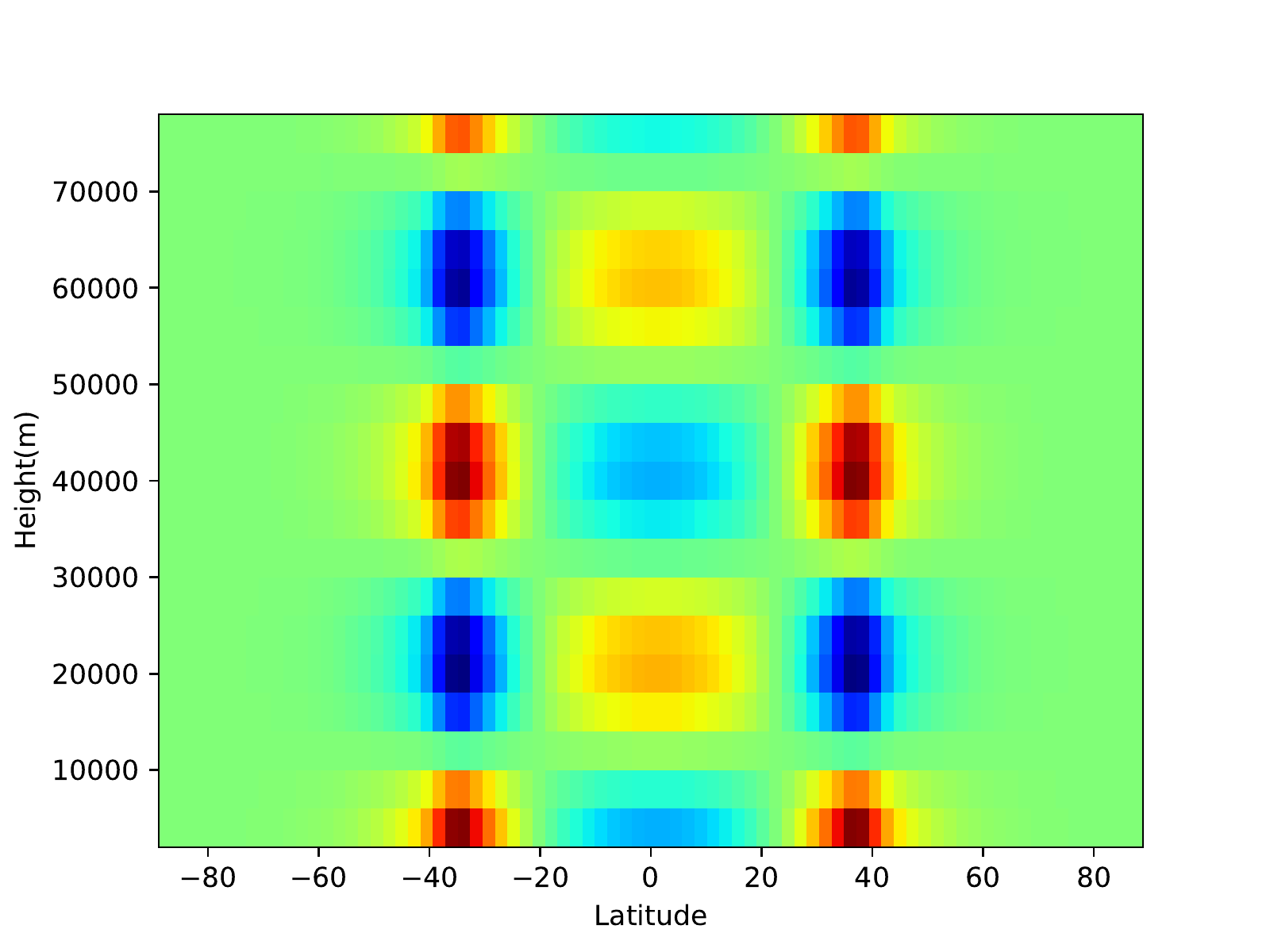}
  \caption{}
  \label{Wang_mitchell_35}
\end{subfigure}%
\begin{subfigure}{.5\textwidth}
  \centering
  \includegraphics[width=\linewidth]{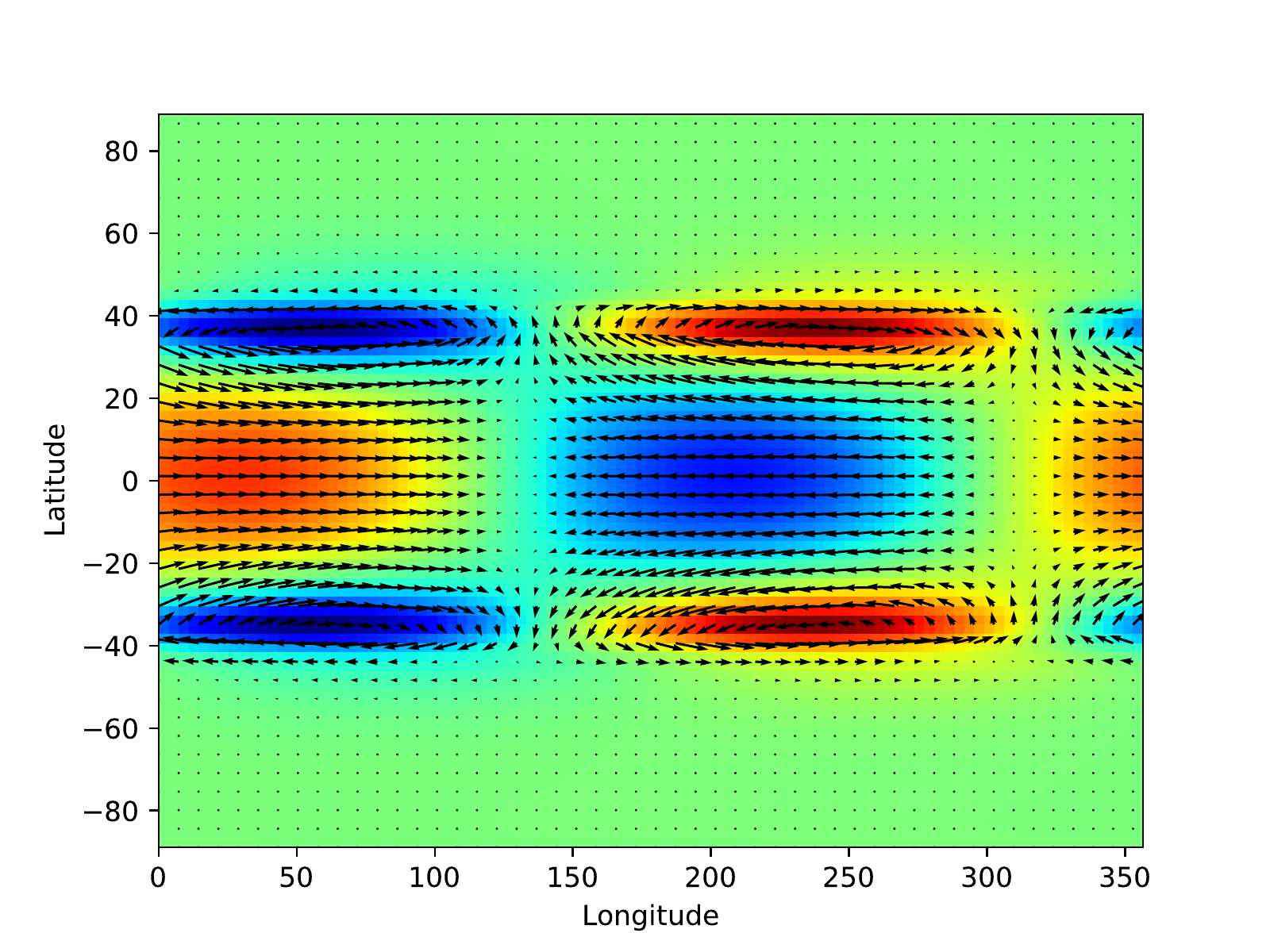}	
  \caption{}
  \label{Wang_mitchell_35_ll}
\end{subfigure}
\caption{Figures showing pressure perturbations (colour scale) and wind vectors (arrows, (b) \& (d) only) for the case of a planet with Earth's radius and rotation rate but an isothermal temperature pressure profile set at $500\,K$.(a) and (c) show latitude against height at a longitude of 200 degrees and (b) and (d) longitude against latitude at a height of 25000 $\mathrm{m}$. We report the values of $\phi_0$ and $Bu$ as defined in \citet{Wang_Mitchell} and the growth rate $\sigma_{growth}$:  (a) and (b): $\phi_0 = 50 ^{\circ}$, $Bu \sim 0.2$ and $\dfrac{\sigma_{growth}}{2 \Omega} = 0.12$, and (c) and (d): $\phi_0 = 35 ^{\circ}$, $Bu \sim 0.05$ and $\dfrac{\sigma_{growth}}{2 \Omega} = 0.026$. These results indicate that the growth rate of the most unstable mode is dependent on the characteristic height of the wave as found in \citet{Wang_Mitchell}.}
\label{fig:Wang_mitchell_compare_2}
\end{figure*}

Following \citet{Wang_Mitchell} we explore the effect on the most unstable mode of varying the planetary parameters. Figure \ref{fig:Wang_mitchell_compare_2} shows results for different $\phi_0$, a characteristic latitude of the initial flow \citep[see][for definitions]{Wang_Mitchell} and the Burger number $Bu = ((N_i H)/(2 \Omega a))^2$ where $H$ is a characteristic height, revealing a change in the growth rate as $H$ is altered \citep[see][for more details]{Wang_Mitchell}. Figure \ref{Wang_mitchell_fig_50} and \ref{Wang_mitchell_fig_50_ll} therefore represent the most unstable mode for a broad inital jet in latitude, up to $50^\circ$, which growth rate is about $\Omega/5$ and characteristic height a third of the total height. On the other hand, Figure \ref{Wang_mitchell_35} and \ref{Wang_mitchell_35_ll} represent the most unstable mode for a narrower inital jet, with $ \phi_0 = 35^\circ$, which growth rate is about $\Omega/20$ and characteristic height a fifth of the total height. The obtained growth rates are consistent with those presented in Figure 2a of \citet{Wang_Mitchell}.

\subsection{Baroclinic instability}
\label{ssec:Ullrich}

\citet{Jablo2006} detail a baroclinic instability test for GCMs using pressure as a vertical coordinate. \citet{Ullrich2014} adapted this test for height--based GCMs. In this test a perturbation to a steady longitudinal wind at mid--latitudes leads to a dynamical instability growing in a few days (for Earth-like conditions). In full 3D GCM simulations many phenomena act simultaneously meaning reproducing an instability with the exact same behaviour and time evolution is unlikely. However, we can expect to reproduce the most unstable modes, which will drive the evolution of the atmosphere in the simulations.

\begin{figure*}[ht!]
\begin{subfigure}{.5\textwidth}
  \centering
  \includegraphics[width=\linewidth]{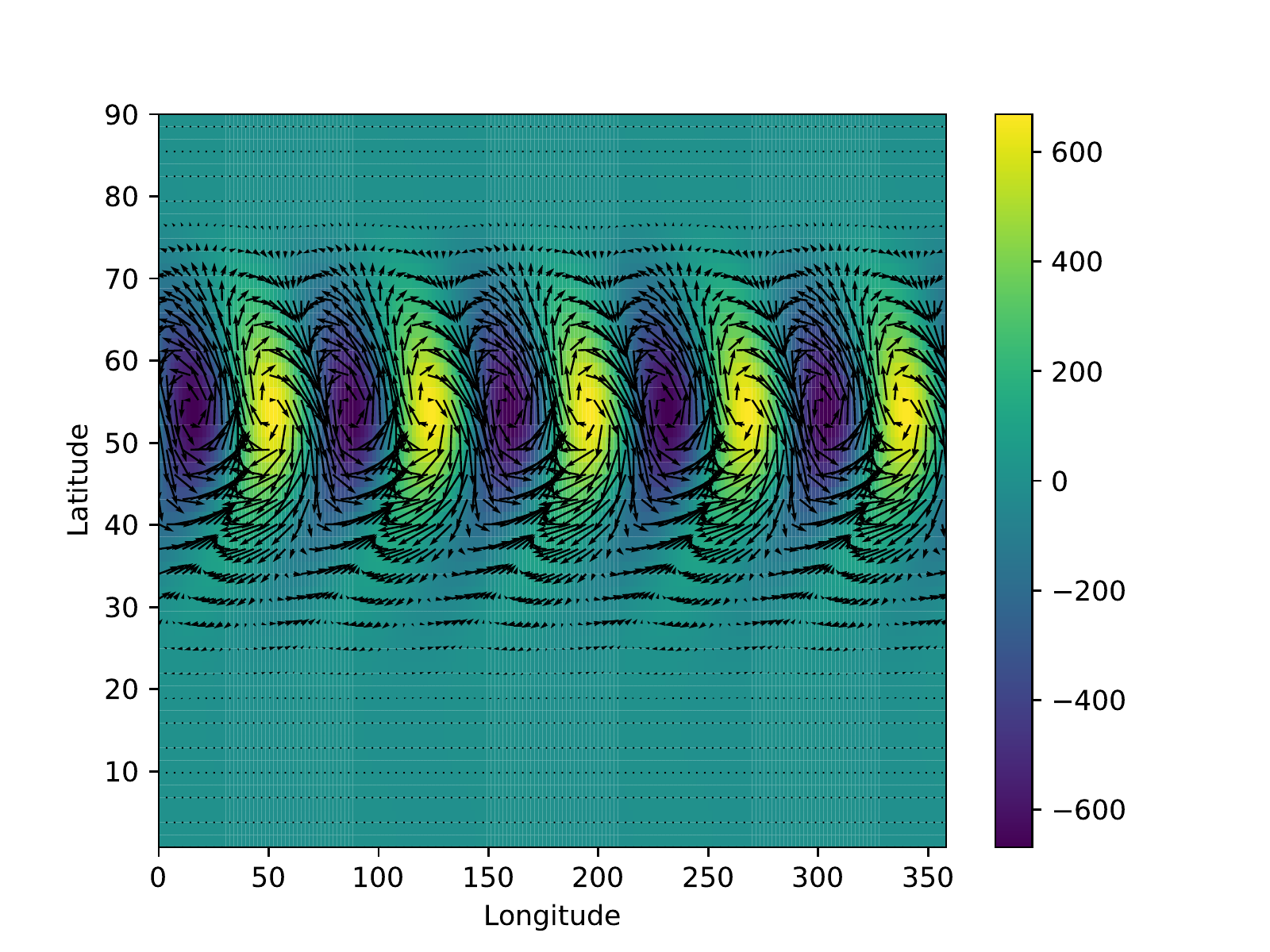}
  \caption{}
  \label{fig:baro_1}
\end{subfigure}%
\begin{subfigure}{.5\textwidth}
  \centering
  \includegraphics[width=\linewidth]{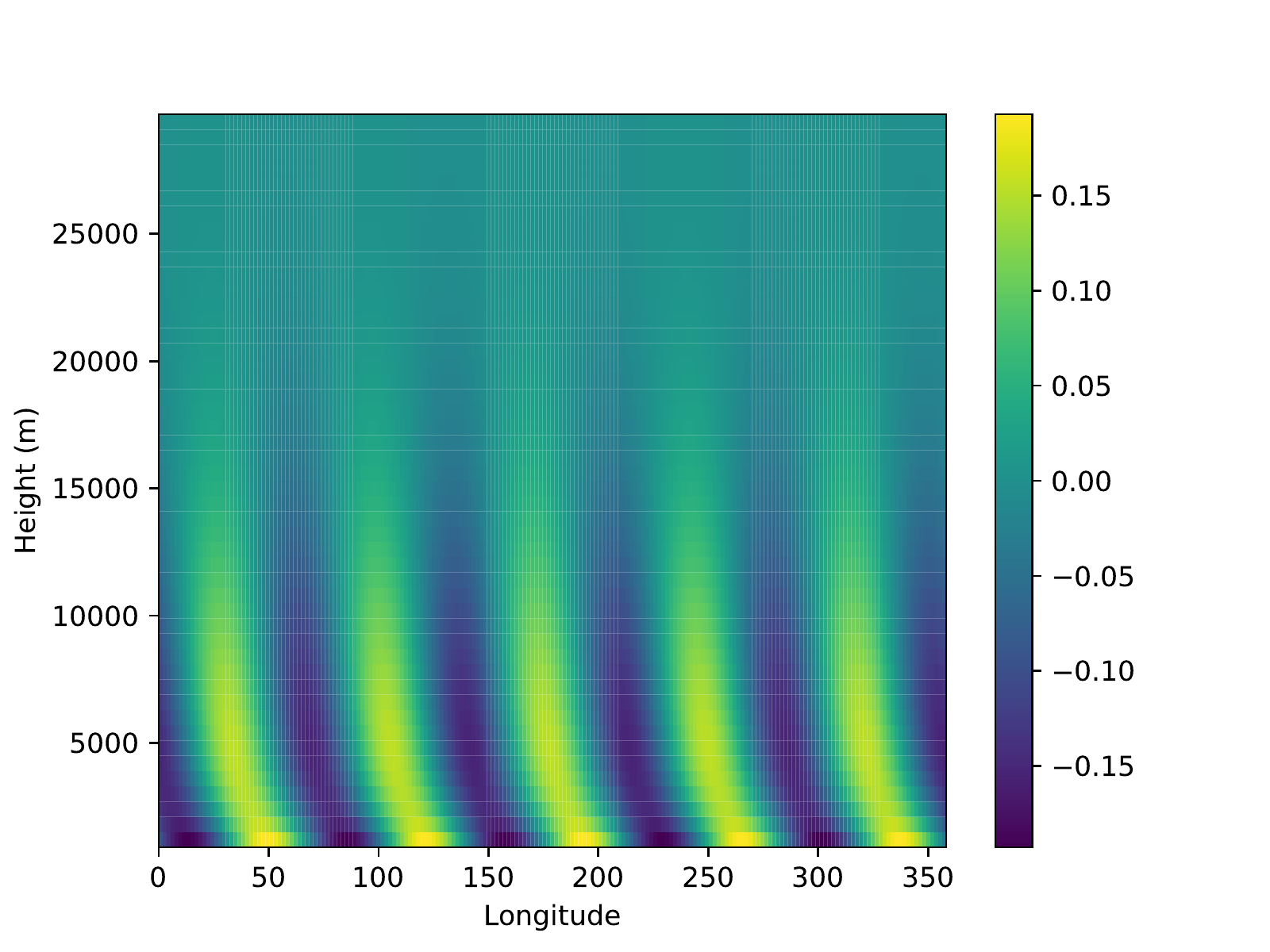}
  \caption{}
  \label{fig:baro_2}
 \end{subfigure}
\caption{Figures showing pressure (colour scale, pascals) and horizontal winds (arrows figure (a) only) using ECLIPS3D, for the baroclinic instability setup of \citet{Ullrich2014}. Figure (a) shows the near--surface pressure as a function of longitude and latitude. Figure (b) then shows pressure as a function of longitude and height at 50$^{\circ}$ latitude}
\label{fig:baro}
\end{figure*}

We have implemented the initial state prescribed in \citet{Ullrich2014} in both the axisymmetric 2D and 3D versions of ECLIPS3D \citep[without the prescribed perturbation of][as ECLIPS3D intrinsically perturbs steady states]{Ullrich2014}. We only show the 2D results as, similarly to the first test case, 2D and 3D are in excellent agreement. In this test, ECLIPS3D identifies the stable modes of \citet{Thuburn2002}, slightly modified by the mean flow and the angular dependency of pressure and temperature. For the unstable modes, a continuum in frequency is returned (discretised by the numerical precision of the algorithm, controlled by the number of points in the matrix $B$), as expected from analytical treatment. ECLIPS3D identifies the most unstable mode at $m=5$ with a growth rate of $6.4 \times 10^{-6} s^{-1}$ ($\sim$2\,days) which is presented in Figure \ref{fig:baro}. Figures 4 and 5 of \citet{Ullrich2014}  demonstrate that the instability dominates the flow after 8\,days, consistent with our growth rate. Additionally, the shape of our instability has similar features to the thermodynamic state of the atmosphere in \citet{Ullrich2014} after 8\, days. Our instability indeed exhibits a tilt in the height vs latitude plot (Figure \ref{fig:baro_2}) as can be seen in Figure 6 of \citet{Ullrich2014}. The pressure also exhibits a similar sharp decrease just above the surface. One must note here that our results include an uncontrolled phase in longitude coming from the axisymmetry of our setup.

The only difference between 2D and 3D is the precision due to discretisation. In 3D, ECLIPS3D is limited as the size of the matrix to invert is much bigger than in 2D. Due to the shape of the staggered grid (Figure \ref{fig:staggered}), the first point in pressure is not the surface pressure and the sharp decrease is less obvious than in 2D (not shown). 

Comparison between the semi--analytical calculation from our a posteriori energy equation and the numerical eigenvalue obtained for this mode reveals close agreement, on the real and imaginary part of $\sigma$, within a few percent. As explained in Appendix \ref{app:energy}, we can decompose this energy equation into three components (five in the case with meridional and vertical velocities) comprising the terms coming from the equations at rest with no angular dependency in the thermodynamic variables, the terms arising from the angular dependency of the steady state and the terms coming from the initial zonal velocity. From analytical considerations, we expect the frequency to be dominated by the velocity terms, where the global mean flow excites the modes at specific phase velocity. However, the growth rate should be controlled by the angular dependent terms as the baroclinic instability
arises from horizontal gradients in the pressure and temperature \citep[see e.g.][]{Vallis06}.

Our energy-based calculation gives a frequency of $1.09 \times 10^{-5} s^{-1}$ where ECLIPS3D finds $9.93 \times 10^{-6} s^{-1}$. In the calculation, the velocity terms account for more than $80 \%$ of the frequency, confirming the analytical predictions. The calculated growth rate is $6.63 \times 10^{-6} s^{-1}$, close to the ECLIPS3D value of $6.39 \times 10^{-6} s^{-1}$, with the angular terms accounting for $96 \%$. These results 
show that our a posteriori energy equation can be a powerful tool to obtain insight in the physics of numerical eigenvectors.

\subsection{Rossby-Haurwitz waves}

The previous steady states we have studied are axisymmetric and include no background velocity (Section \ref{ssec:thuburn}) or only a zonal velocity (Section \ref{ssec:wang} and Section \ref{ssec:Ullrich}). Ideally, we would also benchmark ECLIPS3D using a test including meridional velocities as well as a dependency in longitude. Unfortunately, there are no such non--linear steady states in 3D, in which the analytical theory can provide us with predictions to compare with. We therefore consider a 2D non axisymmetric problem, with steady zonal and meridional winds: Rossby-Haurwitz waves. We here identify the most unstable modes around two steady configuration of this setup which we detail below.

Rossby-Haurwitz waves are analytical 
solutions of the non-linear barotropic vorticity equations. They were discovered by \cite{Haurwitz1940} by perturbing the non--divergent equations and solving them non--linearly. If the flow remains incompressible at all times, these waves remain analytical solutions of the full equations and propagate 
without changing their form at constant speed. With an appropriate choice of parameters, this speed can be 0, and these waves become a stationary, steady solution of the non--divergent barotropic vorticity equations, hence another test case for ECLIPS3D. 

The non-divergent barotropic equations are simply Eq.(\ref{eq:Complete_u}) and Eq.(\ref{eq:Complete_v}) with $w=0$ with an imposed null divergence: 
\begin{equation}
    \dfrac{\partial u}{\partial \lambda} + \dfrac{\partial}{\partial \phi} (v \cos \phi) = 0.
\end{equation}
This last equation does not involve any time derivative, we therefore have to slightly adapt the structure of the code for this set-up. Instead of solving an eigenvector problem, we solve a generalized eigenvector problem where: 
\begin{equation}
    B x = \mathrm{i} \sigma C x,
\end{equation}
where $x$ is an eigenvector, $B$ the linearised matrix of equations and $C$ a diagonal matrix with some zeros in the diagonal. The linearisation
is straightforward, as the equations are similar 
to the full set of equations and the divergence equation is already linear. 

The interest of this test lies in the stability analysis of these waves. \cite{Hoskins1973} showed that Rossby-Haurwitz waves are stable for  longitudinal wave number $R < 5$ and unstable for higher wave numbers. However, in Hoskins' analysis some triad interactions
were missing, as simplifications were required
to treat the problem analytically. Inspired by \cite{Baines1976}, \cite{Thuburn2000} have resolved the 
issue by showing that a Rossby-Haurwitz wave of wavenumber 4 is unstable, because of an interaction with wave numbers 1, 3 and 5
(see also \citet{Lynch2009}). 

ECLIPS3D does not make any assumption on the 
shape of the perturbation needed to trigger an 
instability, nor on the instability itself. We therefore expect to obtain unstable modes around a steady $R=4$ wave. Our set up is 
similar to the classical benchmarking 
test of \cite{Williamson1992} and \cite{Thuburn2000}, 
with a vorticity $\psi$ being:
\begin{equation}
    \psi = - a \omega^2 \sin \phi + a^2 K \cos^R \phi \sin \phi \cos^R \lambda,
\end{equation}
where $K$ and $\omega$ are constants and $R$ is the
longitudinal wavenumber. In order to obtain a steady, stationary wave we also must impose \citep[see][]{Haurwitz1940}:
\begin{equation}
    \dfrac{2R(\Omega + \omega)}{(R+1)(R+2)} = \omega R.
\end{equation}
With $R=4$ and $\Omega = 7.29 \times 10^{-5}\mathrm{s}^{-1}$, the Earth rotation rate,  this leads to $\omega \approx 5 \times 10^{-6} $, close to the value chosen by \cite{Williamson1992} and \cite{Thuburn2000}
$\omega \approx 7.8 \times 10^{-6} $.
With $R=2$, $\omega \approx 1.5 \times 10^{-5}$. In accordance to their set-up, we impose arbitrarily $K = \omega$.

\begin{figure*}[hb!]
\begin{subfigure}{.5\textwidth}
  \centering
  \includegraphics[width=\linewidth]{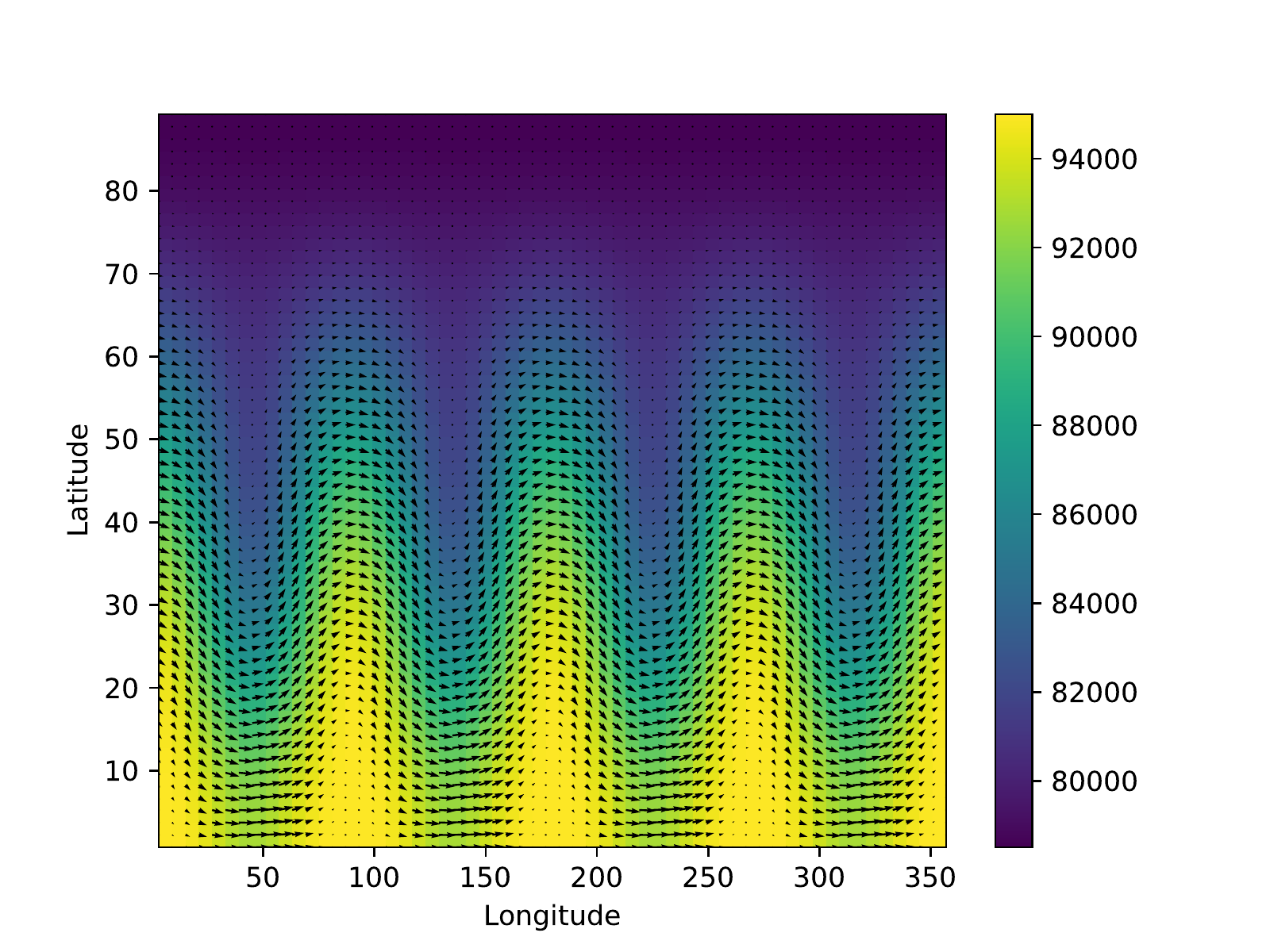}
  \caption{}
  \label{fig:Haurwitz_4}
\end{subfigure}%
\begin{subfigure}{.5\textwidth}
  \centering
  \includegraphics[width=\linewidth]{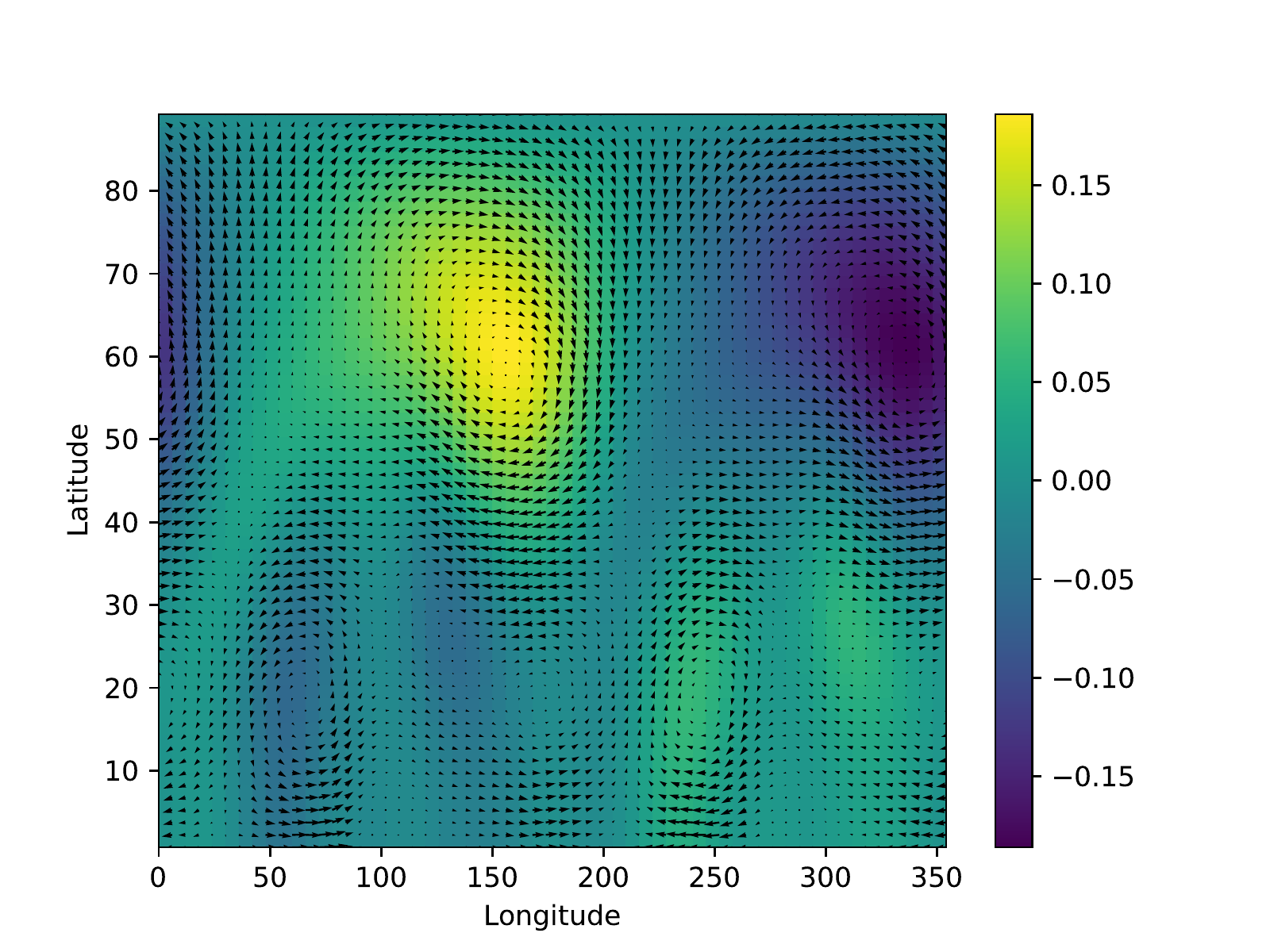}
  \caption{}
  \label{fig:unstable_haurwitz}
\end{subfigure} \\
\begin{subfigure}{.5\textwidth}
  \centering
  \includegraphics[width=\linewidth]{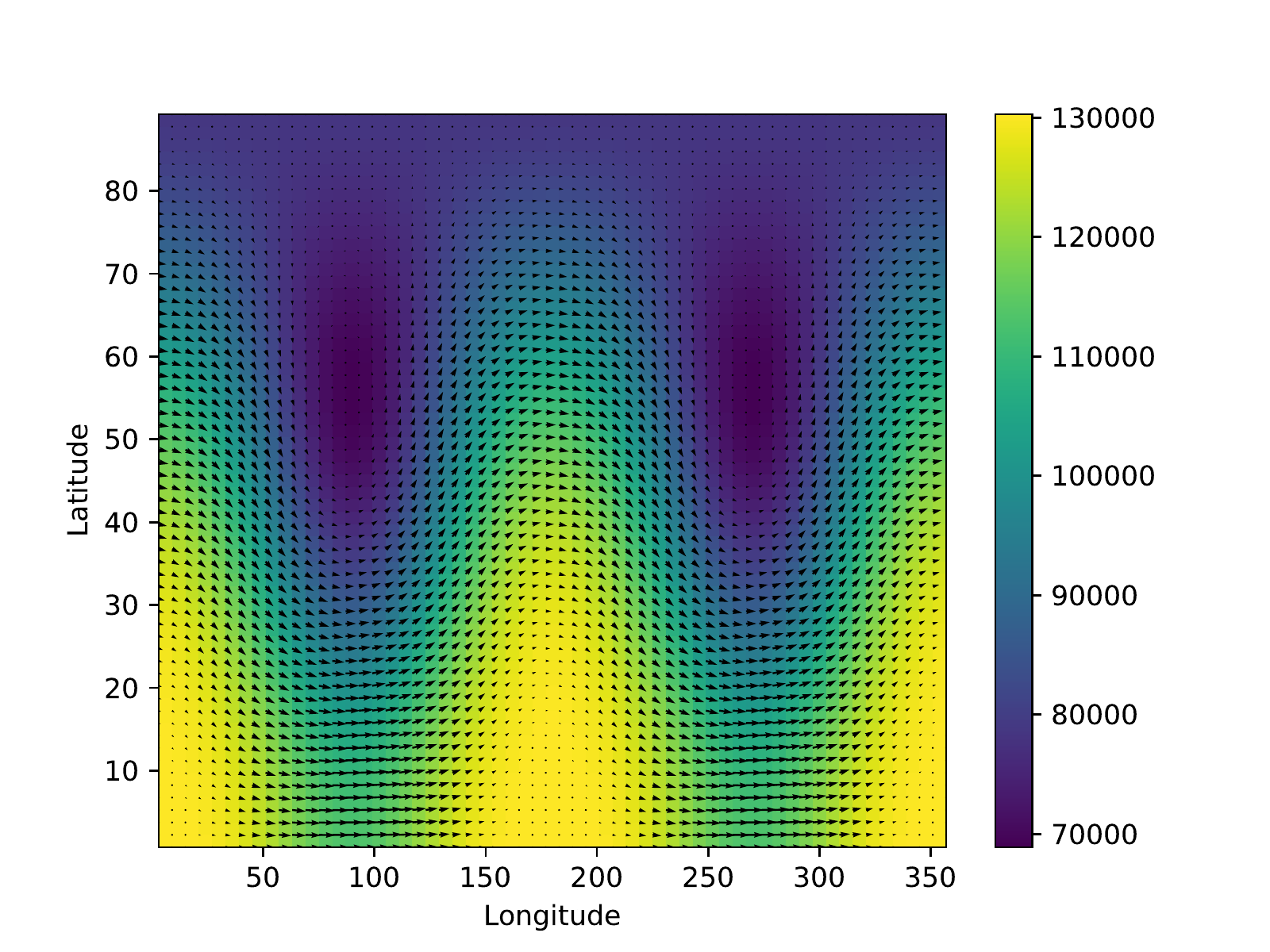}
  \caption{}
  \label{fig:Haurwitz_2}
\end{subfigure}%
\begin{subfigure}{.5\textwidth}
  \centering
  \includegraphics[width=\linewidth]{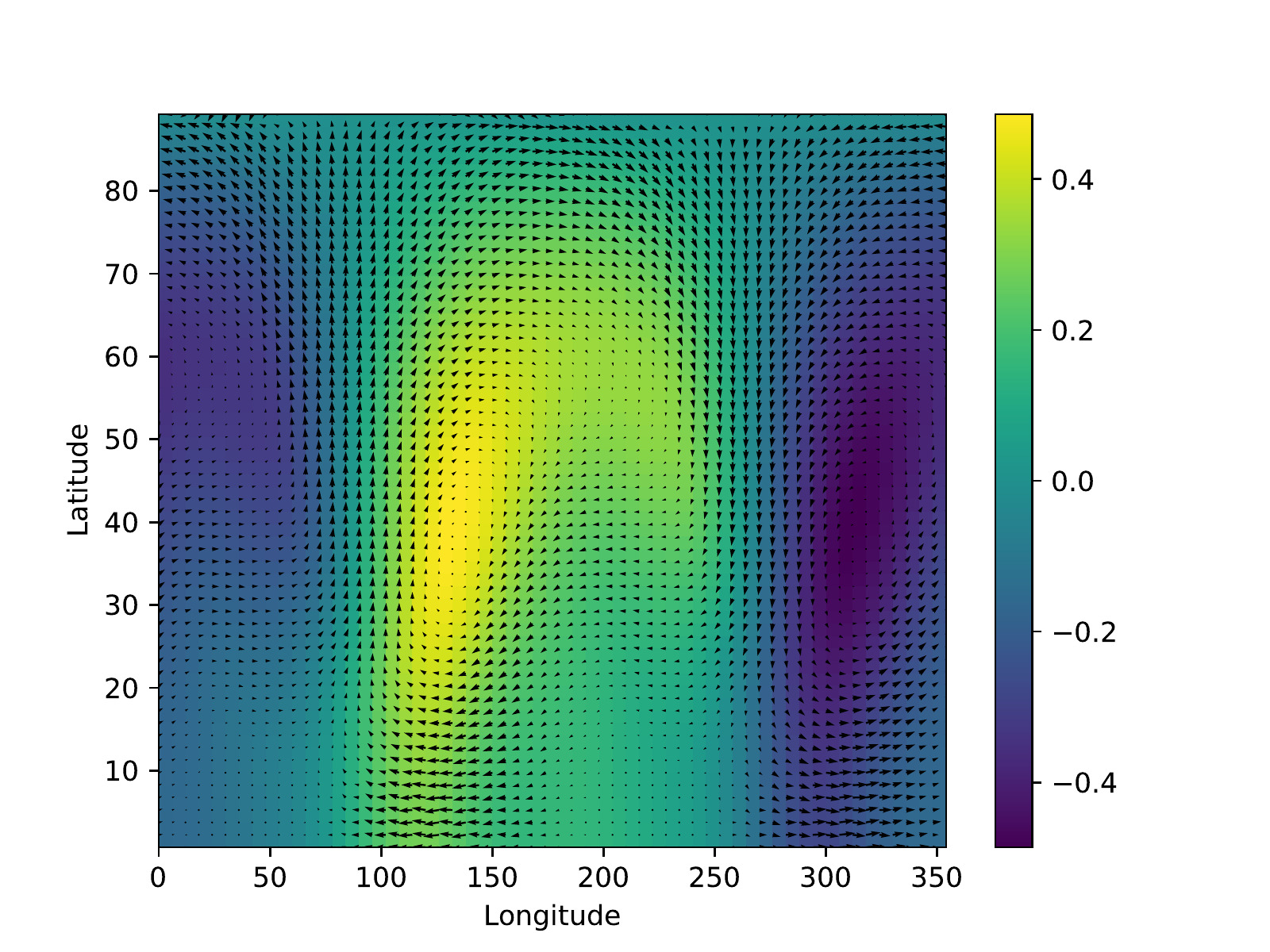}
  \caption{}
  \label{fig:unstable_haurwitz_2}
\end{subfigure}
\caption{Figures showing pressure (colour scale, (a) and (c) in pascals and arbitrary units for (b) and (d)) and horizontal winds (arrows) for Rossby-Haurwitz waves and most unstable modes. Figures (a) and (c) show the initial steady states for $R=4$ and $R=2$ respectively (see text). Figures (b) and (d) are the most unstable mode obtained
with ECLIPS3D for the $R=4$ and $R=2$ set-up respectively.}
\label{fig:haurwitz}
\end{figure*}

We present the initial steady states for $R=4$ and $R=2$ in Figure \ref{fig:Haurwitz_4} and \ref{fig:Haurwitz_2} . As expected from \cite{Thuburn2000}, we obtain a linear instability for the $R=4$ set-up, displayed in Figure \ref{fig:unstable_haurwitz}. This instability oscillates with a period of $\approx 3$ days, with an exponential growth timescale of 6 days. The timescale for instability is globally coherent with the results of \cite{Thuburn2000}, which find that the flow becomes significantly altered
after day 20. For $R=2$, we also find an instability, with striking
resemblance to the $R=4$ instability in shape, a period of just under a day and growth timescale of $4$ days. This mode is shown in Figure \ref{fig:unstable_haurwitz_2}, and this result is in contradiction with \citet{Hoskins1973} but in accordance with \cite{Baines1976} who shows analytically that all the $R \ge 2$ ($n \ge 3$ in their study, where we have $R = n-1$ here) Rossby-Haurwitz waves can be unstable. Interestingly, we also found an instability for a $R=3$ Haurwitz wave but with a growth timescale of more than a hundred days. Such an instability would probably be smoothed out by any source of dissipation or diffusion in a GCM.  Globally, our code agrees well with the theoretical study of Rossby-Haurwitz wave, and demonstates the proper treatment of longitudinal dependent steady state and meridional velocities in ECLIPS3D.

\begin{figure*}[hb!]
\begin{subfigure}{.5\textwidth}
  \centering
  \includegraphics[width=\linewidth]{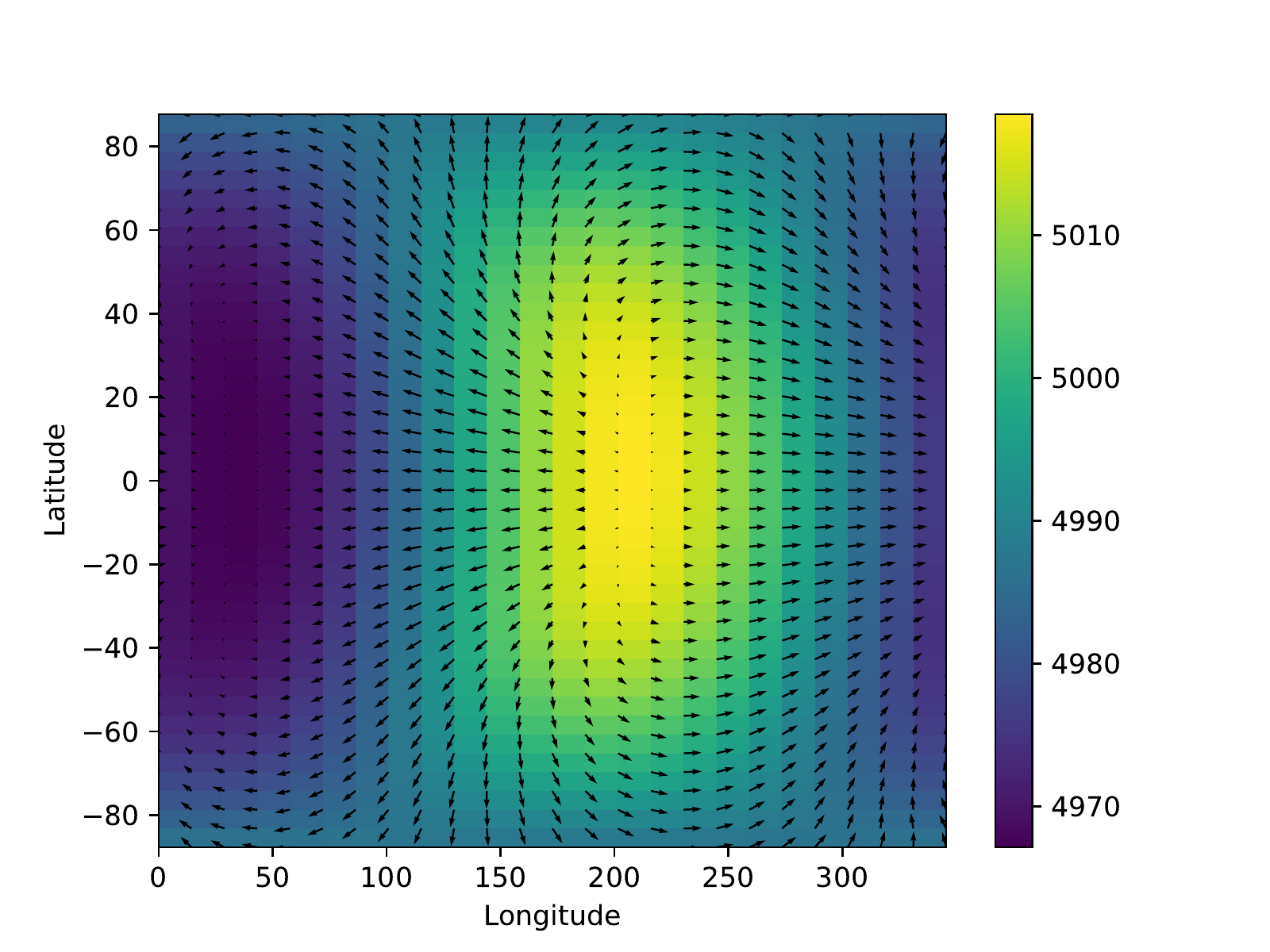}
  \caption{$\tau_\mathrm{drag} = 10^5 s$ and $\tau_\mathrm{rad} = 10^4 s$}
  \label{fig:Komacek_1}
\end{subfigure}%
\begin{subfigure}{.5\textwidth}
  \centering
  \includegraphics[width=\linewidth]{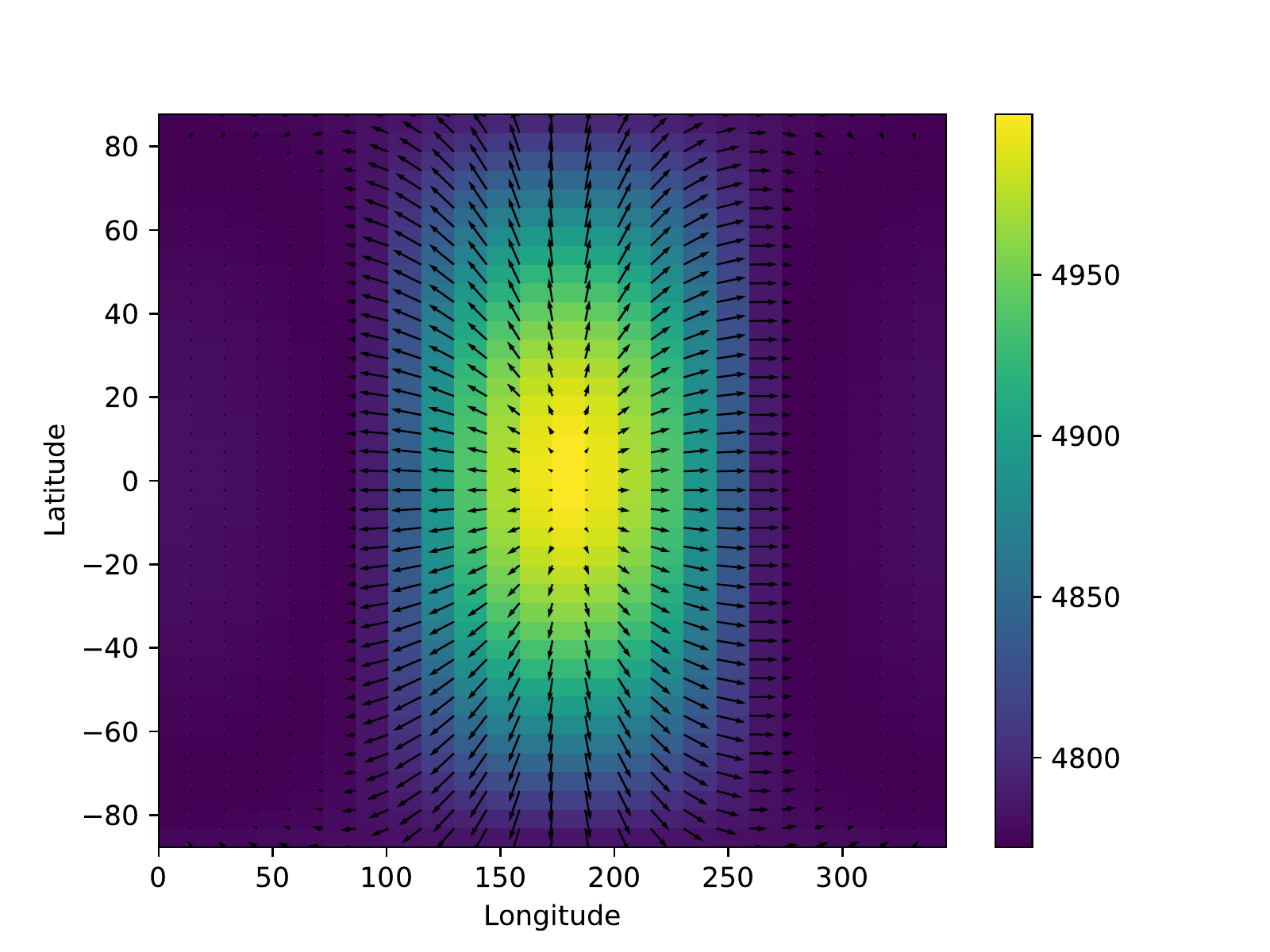}
  \caption{$\tau_\mathrm{drag} = 10^3 s$ and $\tau_\mathrm{rad} = 10^3 s$}
  \label{fig:Komacek_2}
\end{subfigure} \\
\begin{subfigure}{.5\textwidth}
  \centering
  \includegraphics[width=\linewidth]{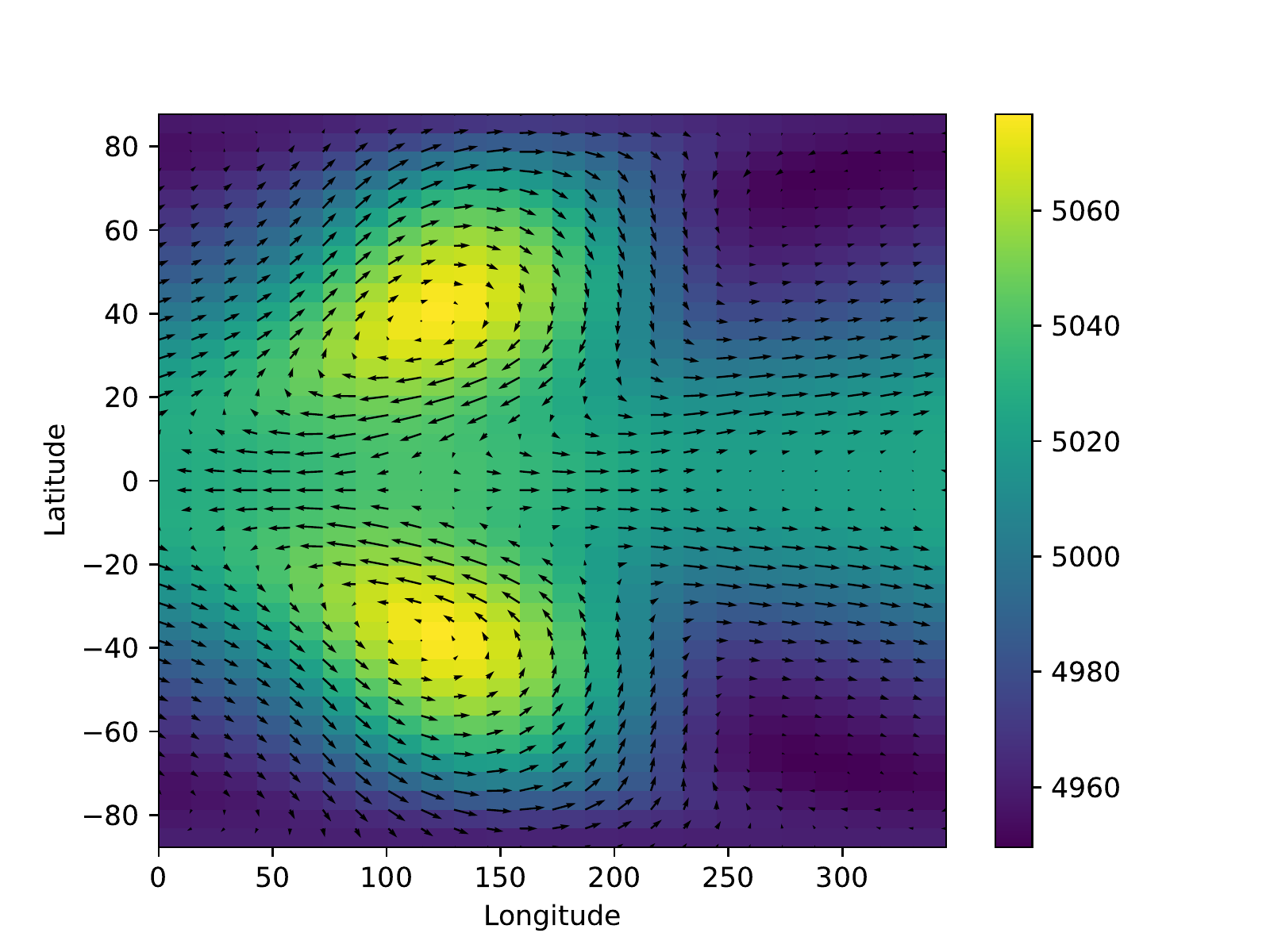}
  \caption{$\tau_\mathrm{drag} = 10^6 s$ and $\tau_\mathrm{rad} = 10^4 s$}
  \label{fig:Komacek_3}
\end{subfigure}%
\begin{subfigure}{.5\textwidth}
  \centering
  \includegraphics[width=\linewidth]{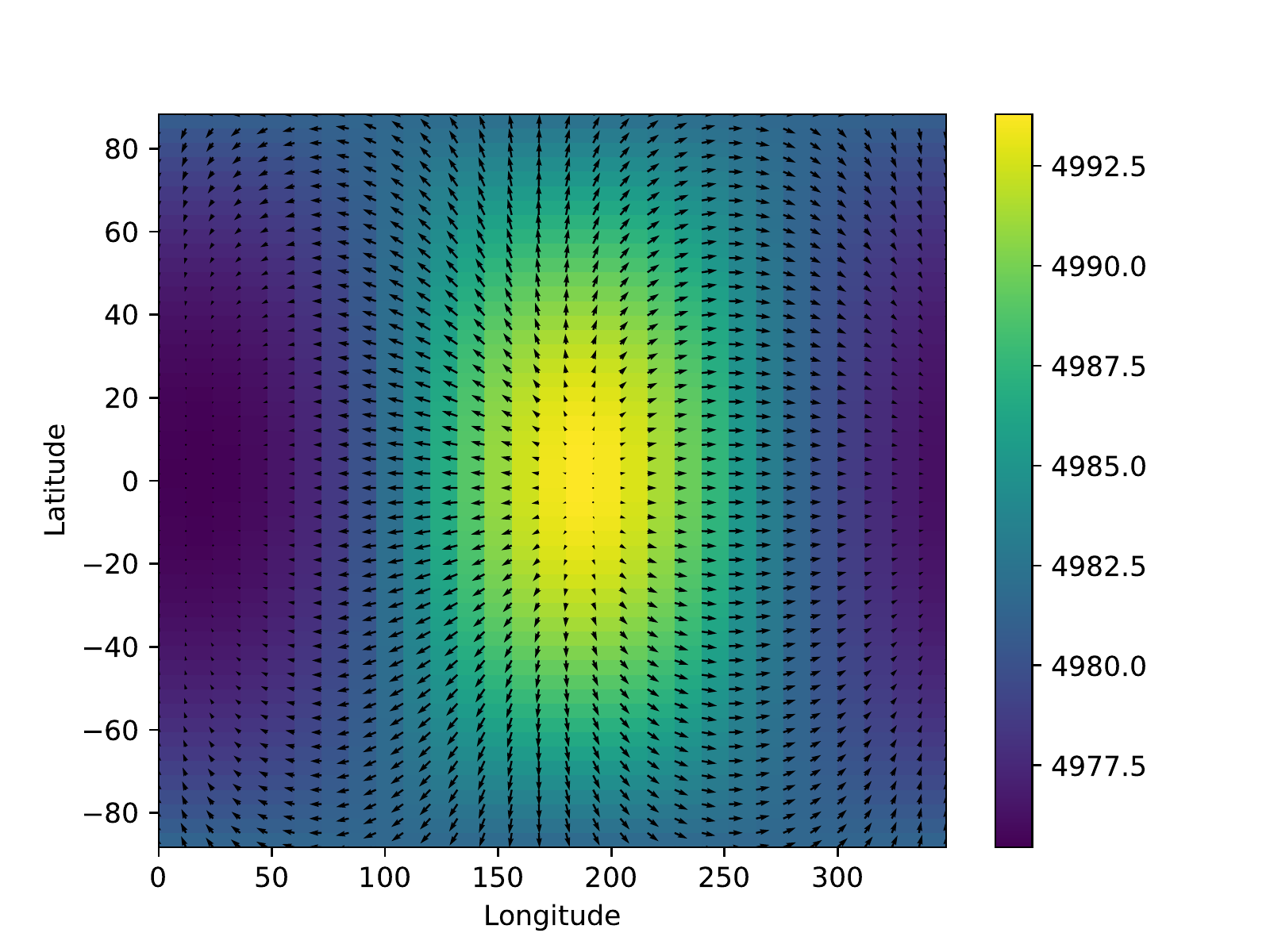}
  \caption{$\tau_\mathrm{drag} = 10^4 s$ and $\tau_\mathrm{rad} = 10^6 s$}
  \label{fig:Komacek_4}
\end{subfigure}
\caption{Figures of the pressure (colour scale, Pascals) and horizontal winds (arrows) as a function of latitude and longitude at a height of $5 \times 10^{6} \mathrm{m}$, for a steady state circulation with a forcing of $\Delta T = 100K$ \citep[see][for definition]{Komacek2016}. Results to be compared to Figure 5 of \citet{Komacek2016}.}
\label{fig:Komacek}
\end{figure*}

\subsection{Linear steady circulation with drag}
\label{ssec:steady}

As discussed in the introduction our development of ECLIPS3D was largely driven by studies of the acceleration of zonal flows in hot Jupiter atmospheres. For these planets analytical studies have shown linear steady states to be of vital importance \citep[see][]{Showman2011}. Therefore, we have also implemented the capability to calculate linear steady states in ECLIPS3D  (see Section \ref{sec:presentation}), which is a much simpler process compared to the identification of eigen modes. To benchmark this section of the code we compare our results to those obtained in the study of \citet{Komacek2016}, in particular the case they present in their Figure 5 where they solve the full Navier-Stokes
equations, but with such a low heating rate that only the linear terms contribute. This requires the addition of a linear drag in the linearised equations following the depth dependent behaviour of that adopted by \citet{Komacek2016}. Additionally, the heating is performed via a Newtonian relaxation with a height--dependent radiative constant. The resulting equation set is shown in Appendix \ref{sapp:steady}. Figure \ref{fig:Komacek} then presents the resulting linear circulations obtained using ECLIPS3D which show excellent qualitative agreement with the results of \citet{Komacek2016} (see their Figure 5). \citet{Komacek2016} do not present the vertical structure of their circulation.

\section{Conclusion}
\label{sec:conclusion}

In this paper we introduced and benchmarked ECLIPS3D: a parallel code for identifying linear instabilities, waves and circulations around a steady state of the Navier--Stokes equations in planetary atmospheres. The linearised equations only omit viscosity and an a posteriori energy equation is used to identify contributions from each component. The time--dependent eigenvector solution or time--independent matrix inversion calculations are performed through discretisation onto a staggered grid and subsequently ScaLAPACK routines.  

The benchmarks cover various well studied wave and instability tests, namely a simple atmosphere at rest \citep{Thuburn2002}, a Rossby-Kelvin unstable jet \citep{Wang_Mitchell}, a baroclinically unstable jet \citep{Ullrich2014}, an unstable Rossby-Haurwitz wave \citep{Thuburn2000} and a linear circulation with atmospheric drag \citep{Komacek2016}. For all these set-ups ECLIPS3D is able to produce excellent,
qualitative agreement with the previous works. We demonstrate that our a posteriori
energy equation is a viable tool to verify the results and identify the dominant terms. We are currently preparing a follow--up study to explore the momentum transfer in hot Jupiter atmospheres and explore the stability of the initial conditions for GCMs using ECLIPS3D (Debras et al., in prep). 

ECLIPS3D currently has several limitations, primarily its computational efficiency, leading to limitations on resolution, particularly for 3D cases. We are  working on several methods to improve this issue for example using libraries adapted to sparse matrices, or splitting the eigenvector solution into several sub--matrices as opposed to a single large matrix (potentially useful as the time taken to solve this type of problem increases faster than linearly with matrix size). This splitting of the matrix may be particularly well suited to a spectral decomposition as we are searching for the most unstable mode, not necessarily trying to capture the entire `shape' of the mode. This could be done through spherical harmonics in the horizontal or Chebyshev's spectral decomposition in the radial direction. 

Despite its limitations, ECLIPS3D in its current version still represents a powerful resource which can be used to study instabilities in 2D situations under axisymmetry or cases where a two--layer model is applicable, or for low resolution 3D problems. The code itself can easily be adapted to different situations, in spherical coordinates, with additional physics or alternative boundary conditions.
As the structure of the code is independent of the underlying equations, meaning alternatives can easily be implemented in terms of symmetries and coordinate systems.

Finally, ECLIPS3D could be applied to a wide range of astrophysical problems. The most obvious one, for which ECLIPS3D was designed, is the study of instabilities and linear circulations for planetary atmospheres, but the range of applicability is greater. Asteroseismology for example requires the need to linearise the equations of motion and identify the leading modes, sometimes with complicated circulation or thermodynamic state inside the star. 
Adapted to cylindrical geometry, ECLIPS3D could be a powerful tool to identify the possible instabilities in protoplanetary disks, where instabilities creating pressure traps are proposed to be a strong way of making planets through core accretion. The addition of a magnetic field in the equations implemented in the code would not pose any theoretical challenge either, which could provide numerous information on the linear behaviour of astrophysical fluids in more general cases.

\begin{acknowledgement}
FD is indebted to the ScaLAPACK team for both the provision of their libraries but also their response to numerous questions. FD thanks the
European Research Council (ERC) for funding under the
H2020 research \& innovation programme (grant agreement
\#740651 NewWorlds). NJM is part funded by a Leverhulme Trust Research Project Grant and partly supported by a Science and Technology Facilities Council Consolidated Grant (ST/R000395/1), both of which we gratefully acknowledge. This study uses material produced using Met Office Software. Additionally, used the DiRAC Complexity system, operated by the University of Leicester IT Services, which forms part of the STFC DiRAC HPC Facility (www.dirac.ac.uk ). This equipment is funded by BIS National E-Infrastructure capital grant ST/K000373/1 and STFC DiRAC Operations grant ST/K0003259/1. DiRAC is part of the National e-Infrastructure. Additionally,  
This research made use of the ISCA High Performance Computing Service at the University of Exeter. Finally, this work was also partly funded by the ERC grant No. 320478-TOFU. 
\end{acknowledgement}

\begin{appendix}
\section{Equations in ECLIPS3D}
\label{app:full_equations}
\subsection{3D, general case}
\label{sapp:3D}
In this appendix we detail the derivation of the full linearised equations for the longitudinal component of the momentum equation, and in the interests of brevity provide only the final expressions for the remaining components. These equations assume a dependence of $g$ on $r$ as $g \propto 1/r^2$, and every other quantity is dependent on the three spatial variables, longitude, latitude and radial distance from the centre of the planet. Here, we consider that $Q$ is $0$ for simplicity, and relax this assumption in \S \ref{sapp:heat}.

The longitudinal equation of momentum is:

\begin{align}
\dfrac{\partial u}{\partial t} + \dfrac{u}{r \cos\phi} \dfrac{\partial u}{\partial \lambda} +
\dfrac{v}{r}\dfrac{\partial u}{\partial \phi} + w \dfrac{\partial u}{\partial r} + 2\Omega w \cos\phi \nonumber \\
- 2 \Omega v \sin\phi + \dfrac{1}{\rho r \cos\phi}\dfrac{\partial p}{\partial \lambda} + \dfrac{uw}{r} - \dfrac{uv \tan\phi}{r}=0.
\end{align}

Term by term we obtain (refer to Eq.\eqref{eq:perturb_var} for definition of perturbed variables): 

\begin{align*}
&\left(\dfrac{\partial u}{\partial t}\right)^{\prime} = \dfrac{1}{\rho_i} \dfrac{\partial u'}{\partial t} \\
&\left(\dfrac{u}{r \cos\phi} \dfrac{\partial u}{\partial \lambda}\right)^{\prime} = \dfrac{u'}{\rho_i r \cos\phi}
 \dfrac{\partial u_i}{\partial \lambda}
\nonumber \\
&+\dfrac{u_i}{r \cos\phi} \left(\dfrac{1}{\rho_i} \dfrac{\partial u'}{\partial \lambda} - \dfrac{u'}{\rho_i^2}
\dfrac{\partial \rho_i}{\partial \lambda}\right) \\
&\left(\dfrac{v}{r} \dfrac{\partial u}{\partial \phi}\right)^{\prime} = \dfrac{v'}{\rho_i r } \dfrac{\partial u_i}{\partial \phi}
+ \dfrac{v_i}{r} \left(\dfrac{1}{\rho_i} \dfrac{\partial u'}{\partial \phi} - \dfrac{u'}{\rho_i^2}
\dfrac{\partial \rho_i}{\partial \phi}\right) \\
&\left(w\dfrac{\partial u}{\partial r}\right)^{\prime} = \dfrac{w'}{\rho_i } \dfrac{\partial u_i}{\partial r}
+ w_i \left(\dfrac{1}{\rho_i} \dfrac{\partial u'}{\partial r} - \dfrac{u'}{\rho_i^2}
\dfrac{\partial \rho_i}{\partial r}\right) \\
&\left(2\Omega w \cos\phi\right)^{\prime} = 2\Omega \dfrac{w'}{\rho_i} \cos\phi \\
&\left(2\Omega v \sin\phi\right)^{\prime} = 2\Omega \dfrac{v'}{\rho_i} \sin\phi \\
&\left(\dfrac{1}{\rho r \cos\phi}\dfrac{\partial p}{\partial \lambda}\right)^{\prime} =
\dfrac{1}{\rho_i r \cos\phi} \left(\dfrac{\partial p'}{\partial \lambda} + \dfrac{1}{\rho_i} 
\dfrac{\partial p_i}{\partial \lambda}\left(\dfrac{\theta'}{g}-\dfrac{p'}{c_i^2}\right)\right)\\
&\left(\dfrac{uw}{r}\right)^{\prime} = \dfrac{1}{\rho_i r} \left(u'w_i+u_iw'\right)\\
&\left(\dfrac{uv\tan\phi}{r}\right)^{\prime} = \dfrac{\tan\phi}{\rho_i r} \left(u'v_i+u_iv'\right).
\end{align*}
Therefore, the final five perturbed equations are:
\begin{align}
&\dfrac{\partial u'}{\partial t} + u' \left(\dfrac{1}{r \cos\phi}
\dfrac{\partial u_i}{\partial \lambda} - \dfrac{u_i}{\rho_i r \cos\phi}\dfrac{\partial \rho_i}{\partial \lambda}
-\dfrac{v_i}{\rho_i r}\dfrac{\partial \rho_i}{\partial \phi} - 
\dfrac{w_i}{\rho_i}\dfrac{\partial \rho_i}{\partial r} \right.\nonumber \\
&\left. + \dfrac{w_i}{r} - \dfrac{v_i \tan\phi}{r}\right) + 
\dfrac{\partial u'}{\partial \lambda} \left(\dfrac{u_i}{r \cos \phi}\right)+ 
\dfrac{\partial u'}{\partial \phi} \left(\dfrac{v_i}{r}\right) + 
\dfrac{\partial u'}{\partial r} \left(w_i\right) \nonumber \\
&+ v' \left(\dfrac{1}{r} \dfrac{\partial u_i}{\partial \phi} - 2 \Omega \sin\phi - \dfrac{u_i \tan\phi}{r}\right) \nonumber \\
&+ w'\left(\dfrac{\partial u_i}{\partial r} + 2 \Omega \cos\phi + \dfrac{u_i}{r}\right)\nonumber \\
&+ p' \left(\dfrac{-1}{c_i^2 \rho_i r \cos\phi} \dfrac{\partial p_i}{\partial \lambda}\right) + 
\dfrac{\partial p'}{\partial \lambda} \left(\dfrac{1}{r \cos\phi}\right) \nonumber \\
& +\theta' \left(\dfrac{1}{g \rho_i r \cos\phi} \dfrac{\partial p_i}{\partial \lambda}\right) = 0 \label{final_u}
\end{align}

\begin{align}
&\dfrac{\partial v'}{\partial t} + u' \left( \dfrac{1}{r \cos \phi} \dfrac{\partial v_i}{\partial \lambda} 
+ 2 \Omega \sin\phi + \dfrac{2 u_i \tan \phi}{r}\right) \nonumber \\
& +v' \left(- \dfrac{u_i}{\rho_i r \cos\phi}\dfrac{\partial \rho_i}{\partial \lambda} + 
\dfrac{1}{r} \dfrac{\partial v_i}{\partial \phi} - \dfrac{v_i}{\rho_i r} \dfrac{\partial \rho_i}{\partial \phi} 
- \dfrac{w_i}{\rho_i} \dfrac{\partial \rho_i}{\partial r} + \dfrac{w_i}{r}\right) \nonumber \\
& +\dfrac{\partial v'}{\partial \lambda} \left(\dfrac{u_i}{r \cos \phi} \right) + 
\dfrac{\partial v'}{\partial \phi} \left(\dfrac{v_i}{r}\right) + \dfrac{\partial v'}{\partial r} \left(w_i\right)
+ w'\left(\dfrac{\partial v_i}{\partial r} + \dfrac{v_i}{r}\right)  \nonumber \\
&+ p' \left(\dfrac{-1}{c_i^2 \rho_i r } \dfrac{\partial p_i}{\partial \phi}\right) + 
\dfrac{\partial p'}{\partial \phi} \left(\dfrac{1}{r}\right) + 
\theta' \left(\dfrac{1}{g \rho_i r} \dfrac{\partial p_i}{\partial \phi}\right) = 0 
\label{final_v}
\end{align}

\begin{align}
&\dfrac{\partial w'}{\partial t} + u' \left(\dfrac{1}{r \cos \phi} \dfrac{\partial w_i}{\partial \lambda} 
- 2 \Omega \cos \phi - \dfrac{2 u_i}{r} \right)  \nonumber \\
&+v' \left( \dfrac{1}{r} \dfrac{\partial w_i}{\partial \phi} -
\dfrac{2 v_i}{r} \right) \nonumber \\
& + w' \left(- \dfrac{u_i}{\rho_i r \cos\phi}\dfrac{\partial \rho_i}{\partial \lambda}
- \dfrac{v_i}{\rho_i r} \dfrac{\partial \rho_i}{\partial \phi} 
- \dfrac{w_i}{\rho_i} \dfrac{\partial \rho_i}{\partial r} + \dfrac{\partial w_i}{\partial r}\right) \nonumber \\
&+\dfrac{\partial w'}{\partial \lambda} \left(\dfrac{u_i}{r \cos \phi} \right) + 
\dfrac{\partial w'}{\partial \phi} \left(\dfrac{v_i}{r}\right) + \dfrac{\partial w'}{\partial r} \left(w_i\right) \nonumber \\
&+ p' \left(\dfrac{-1}{c_i^2 \rho_i} \dfrac{\partial p_i}{\partial r}\right) + 
\dfrac{\partial p'}{\partial r}+ 
\theta' \left(\dfrac{1}{g \rho_i r} \dfrac{\partial p_i}{\partial r}\right) = 0 
\label{final_w}
\end{align}

\begin{align}
&\dfrac{\partial p'}{\partial t} + u' \left(\dfrac{1}{ \rho_i r \cos \phi} \dfrac{\partial p_i}{\partial \lambda} -
\dfrac{c_i^2}{\rho_i r \cos \phi} \dfrac{\partial \rho_i}{\partial \lambda} \right) 
+ \dfrac{\partial u'}{\partial \lambda} \left( \dfrac{c_i^2}{r \cos \phi} \right) \nonumber \\
& + v' \left( \dfrac{1}{ \rho_i r } \dfrac{\partial p_i}{\partial \phi} -
\dfrac{c_i^2}{r} \left(\dfrac{1}{\rho_i}\dfrac{\partial \rho_i}{\partial \phi} + \tan \phi \right) \right)
+ \dfrac{\partial v'}{\partial \phi} \left(\dfrac{c_i^2}{r} \right) \nonumber \\
&+w' \left( c_i^2 \left(\dfrac{2}{r} + \dfrac{N_i^2}{g} \right)\right) 
+ \dfrac{\partial w'}{\partial r} \left(c_i^2 \right) \nonumber \\
&+p' \left(\gamma \left(\dfrac{1}{r \cos \phi}\dfrac{\partial u_i}{\partial \lambda} + 
\dfrac{1}{r}\dfrac{\partial v_i}{\partial \phi} - \dfrac{v_i}{r}\tan \phi 
+ \dfrac{\partial w_i}{\partial r} + \dfrac{2}{r}w_i \right)\right) \nonumber \\
&+ \dfrac{\partial p'}{\partial \lambda} \left(\dfrac{u_i}{r \cos \phi}\right)
+\dfrac{\partial p'}{\partial \phi} \left(\dfrac{v_i}{r}\right) +
\dfrac{\partial p'}{\partial r} \left(w_i \right)= 0
\label{final_p}
\end{align}

\begin{align}
&\dfrac{\partial \theta'}{\partial t} + u' \left(\dfrac{g}{ r \cos \phi} \dfrac{1}{\theta_i}\dfrac{\partial \theta_i}{\partial \lambda}\right) + 
v' \left(\dfrac{g}{ r} \dfrac{1}{\theta_i}\dfrac{\partial \theta_i}{\partial \phi}\right)
+ w' \left(N_i^2\right) \nonumber \\
&+ \theta' \left( \dfrac{u_i}{r \cos \phi} \left(\dfrac{1}{\theta_i}\dfrac{\partial \theta_i}{\partial \lambda} - \dfrac{1}{\rho_i}\dfrac{\partial \rho_i}{\partial \lambda} \right) + 
\dfrac{v_i}{r} \left(\dfrac{1}{\theta_i}\dfrac{\partial \theta_i}{\partial \phi} - \dfrac{1}{\rho_i}\dfrac{\partial \rho_i}{\partial \phi} \right) \right. \nonumber \\
& \left. +w_i  \left(\dfrac{N_i^2}{g} - \dfrac{1}{g}\dfrac{\partial g}{\partial r} -
\dfrac{1}{\rho_i}\dfrac{\partial \rho_i}{\partial r}\right)\right) \nonumber \\
& + \dfrac{\partial \theta'}{\partial \lambda} \left(\dfrac{u_i}{r \cos \phi} \right) + 
\dfrac{\partial \theta'}{\partial \phi} \left(\dfrac{v_i}{r} \right) + 
\dfrac{\partial \theta'}{\partial r} \left(w_i \right) = 0
\label{final_theta}
\end{align}
with $\partial X'/\partial t= i \sigma X'$. \\ 


\subsection{2D, axisymmetric}
\label{sapp:2D}

For the axisymmetric case, the equations are directly obtained from the 3D case 
by choosing a longitudinal wavenumber $m$ such that $X'(t, r, \phi, \lambda) = X'(r,\phi) \mathrm{e}^{\mathrm{i} \sigma t} \mathrm{e}^{\mathrm{i (m/2\pi) \lambda}}$.

\subsection{Two layer equivalent depth}
\label{sapp:twolayer}
The reference for this particular case can be found in \cite{Showman2011} or \cite{Vallis06}.
We consider a dynamic layer above a quiescent layer, reservoir of mass or energy, and study
the horizontal winds $u$ and $v$ as well as the height of the layer $h$, depending on both $x$ and $y$ the
cartesian horizontal coordinates. We follow the definitions of \cite{Showman2011} for the variables, hence consider adimensional equations. There is consequently only three equations to be implemented: 

\begin{align}
    &\dfrac{\partial u'}{\partial t} + u' \left(\dfrac{\partial u_i}{\partial x}\right)
    + \dfrac{\partial u'}{\partial x} \left(u_i\right)
    + \dfrac{\partial u'}{\partial y} \left(v_i\right)  + v'\left( \dfrac{\partial u_i}{\partial y} - y \right) \nonumber \\
    & + \dfrac{\partial h'}{\partial x} = 0
\end{align}

\begin{align}
    &\dfrac{\partial v'}{\partial t} + u' \left(\dfrac{\partial v_i}{\partial x} + y \right)
    + v'\left( \dfrac{\partial v_i}{\partial y}\right)
    + \dfrac{\partial v'}{\partial x} \left(u_i\right)
    + \dfrac{\partial v'}{\partial y} \left(v_i\right)  \nonumber \\
    & + \dfrac{\partial h'}{\partial y} = 0
\end{align}

\begin{align}
    &\dfrac{\partial h'}{\partial t} + u' \left(\dfrac{\partial H}{\partial x}\right) + 
    \dfrac{\partial u'}{\partial x}\left(H\right) +
    v' \left(\dfrac{\partial H}{\partial y}\right) + 
    \dfrac{\partial v'}{\partial y}\left(H\right) + \nonumber \\
    & h' \left(\dfrac{\partial u_i}{\partial x} + \dfrac{\partial v_i}{\partial y}\right) + 
    \dfrac{\partial h'}{\partial x}\left(u_i\right) + 
    \dfrac{\partial h'}{\partial y}\left(v_i\right) = 0  
\end{align}
where $H = H(x,y)$ is the initial steady height.

\subsection{Heating rate}
\label{sapp:heat}

Particular care must be taken when dealing with the heating rate. 
If we call the heating rate $Q_i$, as the initial state is steady the zeroth order term will
cancel the terms involving $Q_i$ in Eq.(\ref{eq:Complete_energy}). However, two situations must be considered: the $\theta/T$ factor in Eq.(\ref{eq:Complete_energy}) has to be linearised, and will be a source of 
additional terms. Additionally,
if $Q_i$ depends on the atmospheric state (for example with Newtonian heating, see next appendix), a perturbation in the atmosphere will be associated with a change $Q'$ in $Q_i$. Therefore, if we write equation Eq.(\ref{final_theta}) as 
\begin{equation}
    \dfrac{g \rho_i}{\theta_i}\left(\dfrac{D \theta}{D t}\right)' = 0 \ \ ,
\end{equation}
where the $g\rho_i/\theta_i$ factor arises from the definition of $\theta'$, the 
final equation involving the heating rate is:
\begin{equation}
    \dfrac{g \rho_i}{\theta_i}\left(\dfrac{D \theta}{D t}\right)' +
    p' \left( \dfrac{g \kappa}{R T_i^2}
    \dfrac{Q_i}{c_p} \right) - \dfrac{Q'}{c_p} \left(\dfrac{g \rho_i}{T_i} \right)= 0,
    \label{eq:theta_Q}
\end{equation}
where $Q'$, if it exists, depends linearly on the linearised atmospheric variables. 

Moreover, obtaining Eq.(\ref{final_p}) implies to use Eq.(\ref{eq:Complete_energy}),
and therefore additional terms also have to be included. More 
precisely, one could show that Eq.(\ref{final_p}) can be written as 
\begin{equation}
    \left(\dfrac{D p}{D t} + \gamma p \vec{\nabla} \cdot \vec{v}\right)' = 0 \ \ .
\end{equation}
With the $Q$ terms we obtain: 
\begin{align}
    &\left(\dfrac{D p}{D t} + \gamma p \vec{\nabla} \cdot \vec{v}\right)' + \gamma R \dfrac{Q_i}{c_p} \left(\dfrac{\theta'}{g} - \dfrac{p'}{c_i^2}\right) + 
    \dfrac{Q'}{c_p} \left(-\gamma R \rho_i\right)= 0 \ \ .
    \label{eq:p_Q}
\end{align}

These new terms in Eqs.(\ref{final_theta}) and (\ref{final_p}) have to be implemented in the matrix from which we solve for eigenvectors, but do not lead to a change in the way of finding the eigenvectors. 

\subsection{Steady linear circulation} 
\label{sapp:steady}

Following \cite{Showman2011} and subsequently \cite{Komacek2016}, we have implemented
the possibility to solve for linear steady states instead of waves and instabilities. We therefore have to impose a heating of the atmosphere, associated to dissipative processes in order to reach a steady state. 

This heating function is extremely different from the heating of Appendix \ref{sapp:heat}: in Appendix \ref{sapp:heat}, we linearized the heating term coming from the initial steady solution of Navier Stokes equations. Here, we prescribe a small forcing of the atmosphere that will make it depart from its initial steady state, and seek for the new steady state that the atmosphere will reach at the linear order (because the heating has a small amplitude). For simplicity reasons, we will consider that the initial steady state was obtained without forcing of the atmosphere (hence $Q$ in Eqs.(\ref{eq:Complete_energy}), (\ref{eq:theta_Q}) and (\ref{eq:p_Q}) is identically null), and call $Q_l$ the small amplitude, linear forcing we impose. 

In that case, 
the perturbed variables are assumed to be constant with time ($\sigma$ is taken to be zero). The dissipative effects will just be linear drags in Eqs.(\ref{eq:Complete_u}), (\ref{eq:Complete_v}) and (\ref{eq:Complete_w}) expressed as $-\vec{v}/\tau_{\mathrm{drag}}$ where $\tau_\mathrm{drag}$ is a characteristic time for the drag, eventually dependent on the space coordinates \citep[see][]{Showman2011}. 

If $Q_l$ is constant, then we just have to modify Eqs.(\ref{final_p}) and (\ref{final_theta}) in a similar way than 
in Appendix \ref{sapp:heat}:

\begin{align}
        &\dfrac{g \rho_i}{\theta_i}\left(\dfrac{D \theta}{D t}\right)' = \dfrac{g \rho_i}{T_i} \dfrac{Q_l}{c_p} \nonumber \\
    &\left(\dfrac{D p}{D t} + \gamma p \vec{\nabla} \cdot \vec{v}\right)' =\gamma R \rho_i\dfrac{Q_l}{c_p} \nonumber
\end{align}
which, as $Q_l$ is order 1, just consisted in neglecting the second order terms in Eqs.(\ref{eq:p_Q}) and (\ref{eq:theta_Q}) and moving the constant heating terms to the right hand side. A dissipative or diffusive process could also be added in the energy equation.

Additionally, a special case must be discussed: Newtonian Heating \citep[see e.g.,][]{Mayne2014}. In that case, $Q_l$ is not constant but depends on the thermodynamic state of the atmosphere. More precisely, calling $Q_\mathrm{N}$ the Newtonian heating rate:
\begin{equation}
    \dfrac{Q_\mathrm{N}}{c_p} = \dfrac{T_\mathrm{eq}-T}{\tau_\mathrm{rad}},
\end{equation}
where $T_\mathrm{eq}$ is a prescribed equilibrium temperature and $\tau_\mathrm{rad}$ a characteristic radiative time, both depending on space variables. 

For the linear forcing approximation to remain correct, $T_\mathrm{eq}$ must be sufficiently close to the initial temperature $T_i$, but then a small change in $T_i$ will have an impact on $Q_\mathrm{N}$ of the same order of $Q_\mathrm{N}$ itself. With our choice of perturbed variables, it is easy to show that:
\begin{equation}
    T' = T_i \left(p' \dfrac{\kappa}{p_i} + \theta' \dfrac{1}{g \rho_i} \right),
\end{equation}
and subsequently 
\begin{equation}
    \dfrac{Q_N}{c_p} = \dfrac{T_\mathrm{eq}-T_i}{\tau_\mathrm{rad}} - \dfrac{T'}{\tau_{\mathrm{rad}}} \equiv \dfrac{Q_{\mathrm{N},i}}{c_p} - \dfrac{T_i}{\tau_\mathrm{rad}} \left(p' \dfrac{\kappa}{p_i} + \theta' \dfrac{1}{g \rho_i}. \right)
\end{equation}
Finally Eqs.(\ref{final_p}) and (\ref{final_theta})
can be rewritten as (using $\gamma R T_i = c_i^2$):
\begin{align}
        &\left(\dfrac{D p}{D t} + \gamma p \vec{\nabla} \cdot \vec{v}\right)' +  \dfrac{c_i^2 \rho_i}{\tau_\mathrm{rad}} \left(p' \dfrac{\kappa}{p_i} + \theta' \dfrac{1}{g \rho_i} \right) = \dfrac{Q_{\mathrm{N},i}}{c_p} \left(\gamma R \rho_i\right)
        \label{eq:steady_p}
        \\
&    \dfrac{g \rho_i}{\theta_i}\left(\dfrac{D \theta}{D t}\right)' + \dfrac{g \rho_i}{\tau_\mathrm{rad}}\left(p' \dfrac{\kappa}{p_i} + \theta' \dfrac{1}{g \rho_i} \right)= \dfrac{Q_{\mathrm{N},i}}{c_p} \left(\dfrac{g \rho_i}{T_i}\right).
\label{eq:steady_theta}
\end{align}
To summarise, when looking for a steady linear circulation with Newtonian heating, we need to invert the matrix $C$ as advertised in Section \ref{ssec:method_nodt}, where $C$ arises from Eqs.(\ref{final_u}) to (\ref{final_theta}) and includes the heating rates and dissipations expressed in the text of this Appendix, and in Eqs.(\ref{eq:steady_p}) and (\ref{eq:steady_theta}).

\section{A posteriori energy equation}
\label{app:energy}

In order to obtain a semi analytical verification for the frequency, we integrate the energy of the modes over the whole volume, and express it as an a posteriori condition on the frequency $\sigma$. In this part, we will assume that the bottom boundary condition is a no escape condition ($w' = 0$) and that the initial state is in the hydrostatic balance: $\dfrac{\partial p_i}{\partial r} = - \rho_i g$, with no initial heating (see \ref{sapp:heat}). These assumptions could be relaxed, but would be sources of numerous additional terms whereas they are always verified in our setups. 

In the 2D axisymmetric case at rest with no angular dependency in the initial variables, \cite{Thuburn2002} used as variables $u'\ , -\mathrm{i} v'\ , -\mathrm{i}w'\ , p'$ and $\theta'$ because this simplifies greatly the calculation. In order to allow for easier verification of our equations, we adopt the same definition for the perturbed variables.
However, for simplicity reason, we drop the primes in the next equation and use $v$ and $w$, not $\mathrm{i} v'$ and $\mathrm{i} w'$. 
Therefore, one has to remember that the $v$ and $w$ expressed in the following equations are actually $-\mathrm{i} v'$ and $-\mathrm{i} w'$ where $v'$ and $w'$ are the solutions of Eq.(\ref{eq:B}). The other variables are not affected.

Denoting a complex conjugate by a star, we express the integral of energy as: 

\begin{align}
    &\iiint_{\Omega} \dfrac{1}{\rho_i}\bigg(\mathrm{i} u^* (\ref{final_u}) + v^* (\ref{final_v}) +  w^* (\ref{final_w}) \nonumber \\
    &+ \mathrm{i} \dfrac{p^*}{c_i^2} (\ref{final_p}) + \mathrm{i} \dfrac{\theta^*}{N_i^2} (\ref{final_theta}) \bigg) \mathrm{d}V = 0
    \label{eq:energy_int_1}
\end{align}
where $\Omega$ is the whole volume, $\mathrm{d}V = r^2 \cos \phi \mathrm{d}r \mathrm{d}\phi \mathrm{d} \lambda$ the infinitesimal volume and (\ref{final_u}) is the left hand side of the complete equation Eq.(\ref{final_u}) etc.

The calculation are really cumbersome, but present no particular difficulty. In order to 
have a physical insight in the leading mechanism from this a posteriori energy equation, we have decided to separate this integral into 5 parts: 
\begin{itemize}
    \item The first part involves only the thermodynamic initial state (no velocities) with a dependency on the radial variable $r$ solely. An initial atmosphere at rest with no angular dependency would have contributions to the energy only from this part.
    \item The second part involves the terms coming from the angular dependency in the thermodynamic steady variables only. 
    \item The third part comes from the steady zonal velocity $u_i$.
    \item The fourth part is generated by the steady meridional velocity $v_i$.
    \item And finally the last part is due to the initial steady vertical velocity $w_i$.
\end{itemize}

Denoting this decomposition of the energy integral as [1] to [5], and 
remembering $\partial /\partial t = - i \sigma$
we obtain an a posteriori equation on $\sigma$ : 

\begin{equation}
    \sigma = - \dfrac{\displaystyle \iiint_{\Omega} \dfrac{1}{\rho_i}\left( [1]+[2]+[3]+[4]+[5] \right) \mathrm{d}V}{\iiint_{\Omega} E \mathrm{d}V}
\end{equation}
This is similar to Eq.(\ref{eq:sigma_energy_int}). The $1/\rho_i$ factor 
might seem useless as it is already in Eq.(\ref{eq:energy_int_1})
but is a necessary density weighting to obtain the appropriate equations. 
$E$ is unchanged:

\begin{equation}
    E = \dfrac{1}{2 \rho_i} \left( (|u|^2 + |v|^2 + |w|^2 + \dfrac{|\theta|^2}{ N^2_i} + \dfrac{|p|^2}{c_i^2} \right)
\end{equation}

After sorting (real component, then imaginary then complex), the calculation gives: 

\begin{flalign}
    &[1] = \Re\bigg(f(u^*v ) - F(u^*w) + \dfrac{\mathrm{i}}{r\cos\phi}u^* \dfrac{\partial p}{\partial \lambda}  \nonumber \\
     &+ \dfrac{1}{r} v^* \dfrac{\partial p}{\partial \phi} + \dfrac{g}{c_i^2} w^* p + w^* \dfrac{\partial p}{\partial r}
    - w^* \theta \bigg) 
\end{flalign}

\begin{flalign}
    &[2] = \Re \bigg[\dfrac{1}{r \cos \phi} \bigg( \dfrac{1}{\rho_i c_i^2} \dfrac{\partial p_i}{\partial \lambda} \mathrm{i} u p^* + \dfrac{\partial p_i/ \partial \lambda}{\partial p_i/ \partial r} \mathrm{i} u \theta^* \bigg)  \nonumber \\
    &- \dfrac{1}{r} \bigg(\dfrac{1}{\rho_i c_i^2} \dfrac{\partial p_i}{\partial \phi}  v p^* + \dfrac{\partial p_i/ \partial \phi}{\partial p_i/ \partial r} v \theta^* \bigg) \bigg] \nonumber \\
    &+ \dfrac{\mathrm{i}}{ 2 r \cos \phi} \bigg(\dfrac{\partial \theta_i/ \partial \lambda}{\partial \theta_i/ \partial r} - \dfrac{\partial p_i/ \partial \lambda}{\partial p_i/ \partial r} \bigg) u \theta^* \nonumber \\
    & - \dfrac{1}{ 2 r}\bigg(\dfrac{\partial \theta_i/ \partial \phi}{\partial \theta_i/ \partial r} - \dfrac{\partial p_i/ \partial \phi}{\partial p_i/ \partial r} \bigg) v \theta^*
\end{flalign}

\begin{flalign}
    &[3] = \Re \bigg(\dfrac{u_i \tan \phi}{r} u^*v - \dfrac{u_i}{r} u^* w \bigg) \nonumber \\
     &+ \dfrac{\mathrm{i}}{2}\bigg[ \dfrac{u_i}{r \cos \phi} \dfrac{\partial \rho_i}{\partial \lambda} \left(- 2 E + \dfrac{1}{\rho_i c_i^2}|p|^2 \right) \nonumber \\
    &+ \dfrac{1}{r \cos \phi}\dfrac{\partial u_i}{\partial \lambda} |u|^2 + 
    \dfrac{u_i}{N_i^2 \theta_i r \cos \phi} \dfrac{\partial \theta_i}{\partial \lambda}|\theta|^2
    \nonumber \\
    & + \dfrac{\gamma}{c_i^2 r \cos \phi} \dfrac{\partial u_i}{\partial \lambda} |p|^2 \bigg]
    \nonumber \\
    & + \dfrac{\mathrm{i}}{2 r \cos \phi} u_i \bigg( u^* \dfrac{\partial u}{\partial \lambda} + 
    v^* \dfrac{\partial v}{\partial \lambda} + w^* \dfrac{\partial w}{\partial \lambda} + 
    \dfrac{\theta^*}{N_i^2} \dfrac{\partial \theta}{\partial \lambda} \nonumber \\
    & + \dfrac{p^*}{c_i^2} \dfrac{\partial p}{\partial \lambda} \bigg) + \dfrac{u_i \tan \phi}{2r} u v^*  - \dfrac{1}{2 r} \dfrac{\partial u_i}{\partial \phi} v u^* \nonumber \\
    &- \dfrac{u_i}{2 r} u w^*
    - \dfrac{1}{2}\dfrac{\partial u_i}{\partial r} w u^*
\end{flalign}

\begin{flalign}
    &[4] = -\Im \bigg(\dfrac{v_i}{r} v^* w\bigg)  \nonumber \\
    & + \dfrac{\mathrm{i}}{2r} \bigg[ v_i \bigg( \dfrac{\partial\rho_i}{\partial \phi} \bigg(-2E + \dfrac{|p|^2}{\rho_i c_i^2}\bigg) - \tan \phi |u|^2 - \dfrac{\gamma \tan \phi}{c_i^2} |p|^2 \nonumber \\
    & + \dfrac{1}{\theta_i} \dfrac{\partial \theta_i}{\partial \phi} \dfrac{|\theta^2|}{N_i^2} \bigg)  +\dfrac{\partial v_i}{\partial \phi} \bigg( |v|^2 + \dfrac{\gamma}{c_i^2} |p|^2 \bigg) \bigg] \nonumber \\
    &+ \mathrm{i}\dfrac{v_i}{2r} \bigg(u^* \dfrac{\partial u}{\partial \phi} + 
    v^* \dfrac{\partial v}{\partial \phi} + w^* \dfrac{\partial w}{\partial \phi} + 
    \dfrac{\theta^*}{N_i^2} \dfrac{\partial \theta}{\partial \phi} + \dfrac{p^*}{c_i^2} \dfrac{\partial p}{\partial \phi}  - w^* v \bigg)\nonumber \\
    & +\dfrac{1}{2 r \cos \phi} \dfrac{\partial v_i}{\partial \lambda} uv^* + \dfrac{\mathrm{i}}{2} \dfrac{\partial v_i}{\partial r} wv^*
\end{flalign}

\begin{align}
    &[5] = \mathrm{i} \dfrac{w_i}{2} \bigg[\dfrac{\partial \rho_i}{\partial r} \bigg(-2E + \dfrac{|p|^2}{\rho_i c_i^2} \bigg) + \dfrac{ |u|^2}{r} + \dfrac{ |v|^2}{r} + \dfrac{ |w|^2}{r} \nonumber \\
    &+\dfrac{|\theta|^2}{N_i^2} \bigg( \dfrac{N_i^2}{g} - \dfrac{1}{g} \dfrac{\partial g}{\partial r} \bigg) + \dfrac{2 \gamma}{c_i^2 r} |p|^2 \bigg] + \dfrac{\mathrm{i}}{2}  \dfrac{\partial w_i}{\partial r} \bigg(|w|^2 +\dfrac{\gamma}{c_i^2} |p|^2 \bigg) \nonumber \\
    &+ \mathrm{i}\dfrac{ w_i}{2} \bigg( u^* \dfrac{\partial u}{\partial r} + 
    v^* \dfrac{\partial v}{\partial r} + w^* \dfrac{\partial w}{\partial r} + 
    \dfrac{\theta^*}{N_i^2} \dfrac{\partial \theta}{\partial r} + \dfrac{p^*}{c_i^2} \dfrac{\partial p}{\partial r} \bigg) \nonumber \\
    &+ \dfrac{1}{ 2 r \cos \phi} \dfrac{\partial w_i}{\partial \lambda} u w^* + \dfrac{\mathrm{i}}{2 r} \dfrac{\partial w_i}{\partial \phi} v w^*
\end{align}

\end{appendix}

\bibliography{bib_benchmark}

\end{document}